\documentclass[]{aa}  

\usepackage{graphicx}
\usepackage{txfonts}
\usepackage{multirow}
\usepackage{graphicx}
\usepackage{grffile}
\usepackage{subfigure}
\usepackage{amsmath}

\usepackage{lineno}
\usepackage{ulem}
\usepackage{soul}
\graphicspath{{./}{figures/}}
\usepackage{natbib}
\usepackage{hyperref}
\begin{document}

   \title{Time-Adaptive PIROCK Method with Error Control for Multi-Fluid and Single-Fluid MHD Systems}

   \subtitle{}

   \author{Q. M. Wargnier
          \inst{1,2}
          \and
          G. Vilmart\inst{3}
          \and J. Mart\'inez-Sykora \inst{1,2,4}
          \and 
          V. H. Hansteen \inst{1,2,4,5} \and
          B. De Pontieu \inst{1,4,5}
          }

   \institute{Lockheed Martin Solar \& Astrophysics Laboratory,
3251 Hanover St, Palo Alto, CA 94304, USA\\
         \and
             Bay Area Environmental Research Institute,
NASA Research Park, Moffett Field, CA 94035, USA\\
\and
Section of Mathematics, University of Geneva, Switzerland \\
\and
Rosseland Centre for Solar Physics, University of Oslo,
P.O. Box 1029 Blindern, NO0315 Oslo, Norway\\
\and 
Institute of Theoretical Astrophysics, University of Oslo, P.O. Box 1029 Blindern, N-0315 Oslo, Norway
             }

   \date{\today}

 
  \abstract
   {The solar atmosphere is a complex environment characterized by numerous species with varying ionization states, particularly evident in the chromosphere, where significant variations in ionization degree occur. This region transitions from highly collisional to weakly collisional states, exhibiting diverse plasma state transitions influenced by varying magnetic strengths and collisional properties. The complexity of processes in the solar atmosphere introduces substantial numerical stiffness in multi-fluid models, leading to severe timestep restrictions in standard time integration methods.}
   {To address the computational challenges, new numerical methods are essential. These methods must effectively manage the diverse timescales associated with multi-fluid and multi-physics models, including convection, dissipative effects, and reactions. The widely used time operator splitting technique provides a straightforward approach but necessitates careful timestep management to prevent stability issues and errors. Despite studies on splitting errors, their impact on solar and stellar astrophysics has largely been overlooked.}
   {We focus on a Multi-Fluid Multi-Species (MFMS) model, which poses significant challenges for time integration. We propose a second-order Partitioned Implicit-Explicit Runge-Kutta (PIROCK) method, introduced by \citet{ABDULLE2013869}. This method combines efficient explicit stabilized and implicit integration techniques while employing variable time-stepping with error control. }
   {Compared to a standard third-order explicit time integration method and a first-order Lie splitting approach as considered by \citet{wargnier2022multifluid}, the PIROCK method demonstrates robust advantages in terms of accuracy, numerical stability, and computational efficiency. For the first time, our results reveal PIROCK's capability to effectively solve multi-fluid problems with unprecedented efficiency. Preliminary results on chemical fractionation, combined with this efficient method, represent a significant step toward understanding the well-known First-Ionization-Potential (FIP) effect in the solar atmosphere.}
   {}


   \maketitle

\newcommand{\eg}{{\it e.g.,}} 
\newcommand{\ie}{{\it i.e.,}} 
\newcommand{\myemail}{wargnier@baeri.org}
\newcommand{\komment}[1]{\texttt{#1}}
\newcommand{\pref}{\protect\ref}
\newcommand{\soho}{{\em SOHO{}}}
\newcommand{\sdo}{{\em SDO{}}}
\newcommand{\stereo}{{\em STEREO{}}}
\newcommand{\iris}{{\em IRIS{}}}
\newcommand{\hinode}{{\em Hinode{}}}
\newcommand{\ebysus}{{\em Ebysus{}}}
\newcommand{\bifrost}{{\em Bifrost{}}}
\newcommand{\wc}{multi-ion cyclotron}
\newcommand{\hp}{H$^{+}$}
\newcommand{\hep}{He$^{+}$}
\newcommand{\cp}{C$^{+}$}
\newcommand{\hepp}{He$^{++}$}
\newcommand{\ms}[1]{\color{brown}{#1}}
\newcommand\msrm{\bgroup\markoverwith{\textcolor{magenta}{\rule[0.5ex]{2pt}{0.4pt}}}\ULon}
\newcommand{\velocity}{\bold u}
\newcommand{\thermalspeed}{\bold u^{\text{th}}} 
\newcommand{\viscous}{\boldsymbol{\tau}} 
\newcommand{\identity}{\mathbf{\mathbb{I}}} 
\newcommand{\EField}{\bold E} 
\newcommand{\BField}{\bold B}
\newcommand{\pvec}{\wedge}
\newcommand{\momex}{\bold R}
\newcommand{\dt}{\partial_t}
\newcommand{\totalcurrent}{\bold J}
\newcommand{\mixture}{\mathcal{M}}
\newcommand{\heavy}{\mathcal{H}}
\newcommand{\elec}{\mathfrak{e}} 
\newcommand{\ions}{\mathcal{I}} 
\newcommand{\Q}{\mathfrak{Q}}
\newcommand{\g}{\mathfrak{g}} 
\newcommand{\ionized}{\mathcal{\hat{I}}} 
\newcommand{\specie}{\mathfrak{a}} 
\newcommand{\fact}[1]{#1\mathpunct{}!}
\newcommand{\crosssection}{C}
\newcommand{\refeq}[1]{Eq.~\ref{#1}}
\newcommand{\refsec}[1]{Section~\ref{#1}}
\newcommand{\reffig}[1]{Fig.~\ref{#1}}
\newcommand{\angstrom}{\mbox{\normalfont\AA}}
\newcommand{\overbar}[1]{\mkern 1.5mu\overline{\mkern-1.5mu#1\mkern-1.5mu}\mkern 1.5mu}

\newcommand{\Fadv}{\bold F_\text{A}}
\newcommand{\Fdiff}{\bold F_\text{D}}
\newcommand{\Freac}{\bold F_\text{R}}
\newcommand{\consvar}{\bold Y}


\section{Introduction}

The solar atmosphere, characterized by complexity, houses numerous species with varying ionization states. This complexity is most obvious in the solar chromosphere which exhibits significant variations in ionization degree, transitioning from highly collisional to weakly collisional states. In particular, the chromosphere exhibits various plasma state transitions, showcasing diverse magnetic strengths and collisional properties across different regions \citep[see][]{Morosin_2022,Przybylski_2022,Martinez_Sykora_2020,Ni_2020}. The growing awareness of the significance of ion-neutral interactions in influencing the dynamics and energy balance of the lower solar atmosphere is evident within the scientific community, in codes developed, and in ongoing efforts \citep[see][and references therein]{Martinez-Sykora:2015fv, Ballester:2018fj, Soler2022}.

When constructing multi-fluid models, the complex nature of processes in the solar atmosphere introduces significant numerical stiffness, leading to severe timestep restrictions based on standard time integration methods. To mitigate computational costs, new numerical methods are required. Attention must be given to the numerical method's capability to handle the diverse timescales associated with multi-fluid and multi-physics models, including convection, dissipative effects, reactions, and other processes. \citet{Alvarez-Laguna:2016ty} employed a numerical strategy based on a full implicit integration to address the numerical stiffness inherent in the multi-fluid MHD equations. However, implicit approaches demand substantial memory resources and involve expensive linear algebra tools. Alternatively, researchers commonly employ conventional explicit schemes, such as Runge-Kutta (RK) methods --- such as the fourth-order RK scheme by \citet{Vogler:2005fj,Navarro_2022,wray2015simulations}. Another notable technique is the third-order 2N Runge-Kutta scheme (RK3-2N) introduced by \citet{carpenterrk32n}, explored by both \citet{Gudiksen:2011qy} and \citet{Pencilcode2021} among others. Despite their ease of implementation, these methods often face challenges in efficiently handling the time integration of stiff equations. The stiffness is primarily attributed to the stringent Courant-Friedrichs-Lewy (CFL) condition they impose, necessitating small timesteps for stability. The impact of this limitation becomes particularly evident in the context of the multi-fluid MHD model \citep[see][]{wargnier2022multifluid,Leake:2012pr,jaraalmonte2021}.

In the field of solar physics, a notable challenge revolves around integrating (in time) diffusive terms like the thermal conduction of electrons and the ambipolar diffusion coefficient. Specifically, codes such as MANCHA3D \citep{Navarro_2022} and MURAM \citep{Vogler:2005fj,Rempel:2017zl} confront this challenge by adopting a parabolic-hyperbolic approach for thermal conduction. This method involves solving a non-Fickian diffusion equation, as elaborated by \citet{Rempel_2016}. An alternative strategy, the Super Time Stepping (STS) method \citep{Genevi96super-time-steppingacceleration} has been widely employed, such as in \citet{Gonzalez_2018} and \citet{Nobreg-Siverio:2020AA...633A..66N} to address ambipolar diffusion, or in \citet{Navarro_2022} to address the thermal conduction. With construction based on Chebyshev polynomials, this STS method can be easily implemented explicitly and allows for a relaxed CFL condition and larger timesteps. However, the main drawback of the STS method is not only its first-order accuracy but also its internal stability issues concerning round-off errors. In the numerical analysis literature, this stability issue was already analyzed in \citet{vanderHouwen1980}, where implementation based on recurrence relations with favorable internal stability has been proposed. Attention is then directed towards the explicit second-order orthogonal Chebyshev method known as ROCK2 \citep{abdulle_2008,Abdulle2002, ROCK2_2001, Zbinden2011}. Although not yet applied in the context of solar physics simulations, the ROCK2 method demonstrates competitiveness with implicit solvers and offers adaptive stability domains tailored for challenging dissipative problems.

Effectively addressing the simultaneous treatment of convective-diffusive-source terms in a multi-fluid MHD model, akin to solving a conventional stiff reaction-convective-diffusion problem, poses a notable challenge in developing a temporal integration method. The commonly used time operator splitting technique, as explored in \citet{strang,duarte,theseduarte,Splitting}, offers a straightforward approach to separately integrating these terms. This method, widely employed in solar physics, as seen in applications like Bifrost \citep{Gudiksen:2011qy}, is suitable for the multi-fluid MHD model through a splitting strategy. Careful management of the timestep is crucial due to potential errors from splitting, which can impact the overall temporal accuracy globally. This holds even when accurate temporal integration methods of high order are considered separately. Attempts to comprehend these errors have been undertaken within applied mathematics, particularly in studies by \citet{descombes_2011} and \citet{Descombes_2014}.  An alternative strategy involves partitioned implicit-explicit Runge-Kutta methods (IRKC), as proposed by \citet{Verwer2004}. Additionally, a second-order partitioned method, PRKC, is discussed by \citet{Zbinden2011} combining a second-order explicit stabilized method with a third-order explicit Runge-Kutta method. In this work, we have considered the PIROCK method developed by \citet{ABDULLE2013869}, a partitioned Runge-Kutta method inspired by the earlier approaches
\citep{Abdulle2002, Verwer2004, Zbinden2011} and combining efficient integration methods, precisely ROCK2 for diffusive terms, a third-order Runge-Kutta Heun method, and a second-order Singly Diagonally Implicit Runge-Kutta (SDIRK) method for reactive terms. The method incorporates an adaptive timestep strategy using local error estimators based on embedded methods.

We consider a comprehensive Multi-Fluid Multi-Species (MFMS) model detailed in our prior work \citet{wargnier2022multifluid}, addressing a convective-diffusive-reaction problem with the inclusion of the Hall term. In Section~\ref{sec:model}, we provide a brief overview of the model within the context of a single ordinary differential equation. Following that, we detail the spatial discretization and the formulation of hyperdiffusive terms incorporated into the model in Section~\ref{sec:hyp}. Moving on to Section~\ref{sec:rkmethods}, we review methods found in the literature for separately integrating diffusive, advective, and stiff reactive terms, using a simple Dahlquist test problem as a reference. We also discuss the time integration strategy employed in the solar physics literature for integrating these terms collectively. In Section~\ref{sec:numerical}, we provide a brief description of the PIROCK algorithm, its modification, and the adaptive timestep control strategy based on error estimators. In Section~\ref{sec:tests}, we present and compare simulation results using PIROCK against third-order explicit time integration and a first-order Lie splitting method \citep[e.g.,][]{wargnier2022multifluid}. We aim to demonstrate PIROCK as the optimal strategy for MFMS model time integration, significantly reducing computational costs while maintaining second-order accuracy. We highlight its efficiency compared to classical explicit methods, such as the Bifrost code for single-fluid MHD models, including hyperdiffusive terms and ambipolar diffusion.

\section{Multi-fluid multi-species governing equations}~\label{sec:model}
This section provides an overview of the Multi-fluid Multi-Species (MFMS) model. The model was previously described in Section~2 of \citet{wargnier2022multifluid}, where additional details were provided. In subsequent discourse, the notation $i$ shall represent a fluid of species associated with a heavy particle at any ionization level or excited state, excluding electrons, within the composite mixture of electrons and heavy particles. The notation $\elec$ represents the electron fluid. 

In contrast to previous work \citep{wargnier2022multifluid}, we account for all terms without any simplifications, including the Hall term. After discretization in space, the system of equations can be straightforwardly reformulated as a set of ordinary differential equations that are segregated into different terms as depicted by the equation:

\begin{equation}
\frac{d\consvar(t)}{dt} = \Fadv\left(\consvar(t)\right) + \Freac\left(\consvar(t)\right) + \Fdiff\left(\consvar(t)\right) ,
\label{eq:system}
\end{equation}
where $\consvar = \left(\rho_{i},\;\rho_{i} \velocity_{i}^T,\; e_{i},\; e_{\elec},\;\BField^T \right)^T$ is the vector of conservative variables that involves the mass density $\rho_{i}$, velocity $\velocity_{i}$, thermal energies $e_{i}$ of each ionized/excited or ground level heavy species, as well as the electron thermal energy $e_{\elec}$ and the magnetic field $\BField$.  $\Fadv$, $\Fdiff$, and $\Freac$ are the advective, diffusive fluxes and reactive terms, respectively, and are defined in a compact form as follows

	\begin{equation}
    \left\{
    \begin{aligned}
        \Fadv(\consvar ) &=-\nabla\cdot\bold f_{\text{A}}(\consvar )-\bold S_{\text{A}}(\consvar,\nabla\consvar ), \\
        \bold f_{\text{A}}(\consvar ) &= \left(\rho_{i} \velocity_{i},\;\rho_{i} \velocity_{i}\otimes\velocity_{i} + P_{i}\identity,\; e_{i}\velocity_{i},\; e_{\elec}\velocity_{\elec},\;\identity\pvec\left(\velocity_{\elec}\pvec\BField\right)\right)^T, \\
        \bold S_{\text{A}}(\consvar,\nabla\consvar  ) &= \Biggl(0_i,\;n_i q_i \left( \left[\velocity_{\elec}-\velocity_i\right] \pvec \BField -\frac{\nabla P_{\elec}}{n_{\elec} q_{\elec}} \right),\\
        & \quad\quad\quad\quad P_i\nabla\cdot\velocity_i,\;P_\elec\nabla\cdot\velocity_\elec,\;-\nabla\pvec\left(\frac{\nabla P_{\elec}}{n_{\elec}q_{\elec}}\right) \Biggr)^T
    \end{aligned}
    \right.
\end{equation}
	\begin{equation}
		\left\{
	\\
	\begin{aligned}
	\Fdiff(\consvar ) &=\nabla\cdot \bold f_{\text{D}}(\nabla\consvar )  \\
	\bold f_{\text{D}}(\nabla\consvar) &= \left(0_i,\;\viscous^{*}_{i},\;-\viscous^{*}_{i} : \nabla\velocity_i ,\; 
 H^{\text{spitz}}_{\elec},\;\identity\pvec\left(\frac{\momex_\mathrm{\elec}^{\text{col}}}{n_\elec q_\elec}+\EField^{*}\right) \right)^T
		\end{aligned}
	\right.
	\end{equation}
    \begin{equation}
    \left\{
	\\
	\begin{aligned}
	\Freac\left(\consvar\right) &=\biggl(\Gamma^{\text{ion,rec}}_{i},\;\momex^{\text{ion,rec}}_i +\momex_i^{\text{col}} +\left[\frac{n_{i}q_{i}}{n_{\elec}q_{\elec}}\right]\momex_\mathrm{\elec}^{\text{col}},\; \\&\quad\quad\quad\quad H^{\text{ion,rec}}_i +          H_i^{\text{col}}, \;H^{\text{ion,rec}}_\mathrm{\elec} +   H_\mathrm{\elec}^{\text{col}},\;0_3\biggr)^T\nonumber
		\end{aligned}
  \right.
	\end{equation}
where $P_{i}$ and $P_{\elec}$ denote the pressure of a particular fluid species $i$, and the electron pressure respectively. $\identity$ is the identity matrix, $0_3$ is the null matrices of size three, $n_{i}$ and $n_{\elec}$ are the number densities of particular fluid species $i$ and electrons respectively, while $q_{i}$ and $q_{\elec}$ are the ion and electron charges, respectively. 

The advective terms $\Fadv$ are split into advective fluxes $\nabla\cdot \bold f_{\text{A}}(\consvar )$ and source terms $\bold S_{\text{A}}(\consvar,\nabla\consvar)$. In contrast to the reactive terms $\Freac$ discussed later, the advection source terms $\bold S_{\text{A}}$ have the particularity of depending on the gradient of the conservative variables $\nabla \consvar$, rendering them less suitable for integration with an implicit method, as considered for $\Freac$. Furthermore, $\bold S_{\text{A}}$ includes non-conservative products like $P_i\nabla\cdot\velocity_i$ or $P_{\elec}\nabla\cdot\velocity_{\elec}$. The source term $\bold S_{\text{A}}$ includes also the battery term $\nabla P_{\elec}/(n_{\elec} q_{\elec})$ and the ion-coupling term $n_i q_i \left( \left[\velocity_{\elec}-\velocity_i\right] \pvec \BField\right)$. In our study, we have also treated the ion-coupling term as a reactive source term $\Freac$, assuming a constant magnetic field and current during its integration. This strategy is convenient in cases where the term becomes numerically stiff. Our results, which demonstrate satisfactory performance in calculations and accuracy, support the feasibility of this approach.

The diffusive fluxes $\Fdiff$ encompass physical diffusive terms, such as $H^{\text{spitz}}_{\elec}$ representing the anisotropic electron heat flux, and $\momex_\mathrm{\elec}^{\text{col}}/({n_\elec q_\elec})$, which denotes the collision term derived from the electron momentum equation and is defined within the electric field formulation. Additionally, artificial hyperdiffusive terms such as $\viscous^{*}_{i}$ and $\EField^{*}$ for each momentum equation associated with species $i$ and magnetic field are included. Further elaboration on the hyperdiffusive terms in the MFMS context is provided in the subsequent section~\ref{sec:hyp}. The diffusive term, being the divergence of a function dependent on the gradient of the conservative variable, behaves akin to a second-order differential operator.

In the definition of the reactive terms $\Freac$, $\Gamma^{\text{ion},\text{rec}}_{i}$ is the sum of the mass transition rate due to recombination or de-excitation, and ionization or excitation, associated with fluid species $i$. $\momex^{\text{ion},\text{rec}}_i$ is sum of all the momentum exchange terms for species $i$ due to both ionization and recombination processes. $H^{\text{ion},\text{rec}}_{i}$ and $H^{\text{ion},\text{rec}}_{\elec}$ are the sum of the heating and cooling terms due to the ionization and recombination processes associated with particles $i$ and electrons, respectively. $\momex_{i}^{\text{col}}$ is the sum of all the momentum exchange terms due to collisions between fluid species $i$ with any other fluid species in the mixture considered, $H_i^{\text{col}}$ is the sum of all the heating terms due to collisions associated with particle $i$.  Similarly, as in \citet{wargnier2022multifluid}, the radiative losses are optically thin in the term $H^{\text{ion},\text{rec}}_{\elec}$. Unlike advective or diffusive terms, these reactive terms lack spatial connectivity. Their numerical stiffness arises from their association with small timescales, such as collisions, ionization, and recombination processes.

Given the quasi-neutrality assumption and the presence of multiple ionized species, the electron velocity can be expressed as a function of the hydrodynamic velocity of each ion and the total current. This is written as:

\begin{equation}
    \velocity_{\elec} = \sum_{i}\frac{n_{i}q_{i}\velocity_{i}}{n_{\elec}q_{\elec}}-\frac{\totalcurrent}{n_{\elec}q_{\elec}},
    \label{eq:vele}
\end{equation}
where $\totalcurrent=(\nabla \pvec \BField)/\mu_0$ is the total current defined from Maxwell-Ampere's law and $\mu_0$ is the vacuum permeability. Note that the Hall term appears in the magnetic induction, \ie\, in $f_{\text{A}}$ by considering the contribution of the total current of the electron velocity in the second term of Eq.~\eqref{eq:vele}.

\section{Spatial discretization and hyper-diffusive terms treated as $\Fdiff$}
\label{sec:hyp}

We emphasize that this paper mainly focuses on temporal integration while the considered spatial discretization method utilized in this study closely resembles the approach outlined by \citet{Gudiksen:2011qy}, as detailed in the previous work \citet{wargnier2022multifluid}. It entails the application of a sixth-order differential operator for calculating gradients associated with convective and diffusive fluxes,  along with a fifth-order interpolation scheme for repositioning conservative quantities on a staggered grid. We highlight that the method presented here is general and could be adapted straightforwardly to other choices of spatial discretization methods. After spatial discretization, it treats temporal integration as an ordinary differential equation, as depicted in Eq.~\eqref{eq:system}. 

The high-order finite difference scheme necessitates the incorporation of artificial numerical terms, referred to as hyperdiffusive terms. Analogous to \citet{Nordlund1982} in the single-fluid MHD case, these terms are employed for stability and mitigate spurious oscillations near shocks or discontinuities, also known as the Gibbs phenomenon. The initial MFMS model is adjusted for numerical approximation needs by including artificial hyperdiffusive terms to enhance stability. It is important to note that these terms are treated as diffusive terms ($\Fdiff$). They will be temporarily integrated using the same method employed for physical diffusive terms, such as electron thermal conduction.

The hyperdiffusive terms are defined by $\viscous^{*}_{i}$ and $\EField^{*}$ in $\bold f_{\text{D}}$. The artificial viscous stress tensor $\viscous^{*}_{i}$ for each momentum equation is defined in a parabolic form for any fluid-species $i$, as follows:

\begin{equation}
\nabla\cdot\viscous^{*}_{i} = \nabla\cdot \left(\alpha_i(\velocity_i)\nabla(\velocity_i)\right)
\end{equation}

\begin{equation}
    \alpha_i(\velocity_i) = \left( C^k_i\, \Delta k\, Q(\partial_k \text{u}_{i,j}) \right)_{k,j\in\{x,y,z\}^2}
\label{eq:hypdiff}
 \end{equation}
where $\Delta k$ denote the grid spacing sizes along dimensions $x$, $y$, or $z$, $Q$ defined by \citet{Gudiksen:2011qy} functions as a quenching operator aimed at detecting strong gradients or discontinuities in space and amplifying diffusion within those regions, $\partial_k\text{u}_{i,j}$ denotes the spatial derivative of the $j$-component of velocity $\velocity_i$ along dimension $k$, $C^k_{i}$ is a factor defined as :
\begin{equation}
C^k_{i} = \rho_i\left(\nu_1 c_{i,f} + \nu_2 |\text{u}_{i,k}| +\nu_3 \Delta k |\partial_k\text{u}_{i,k}|\right),
\label{eq:ccoeff}
\end{equation}
where $\nu_1$,  $\nu_2$ and $\nu_3$ are arbitrary coefficients typically ranging between 0 and 1, $c_{i,f}$ is the fast magnetosonic wave associated with species $i$.

Our study incorporates artificial hyperdiffusive terms into the magnetic field equation represented by $\EField^*$. This approach mirrors that of \citet{Nordlund1982} for single-fluid MHD but has been extended to MFMS, analogous to the treatment of the viscous stress tensor discussed previously. We introduce supplementary artificial hyperdiffusive terms characterized by timescales corresponding to the Hall and ion-coupling terms, which are absent in the conventional single-fluid MHD framework. Additional elucidation regarding these adjustments aligns with the methodology outlined by \citet{Nobrega2020}  within the framework of the ambipolar diffusion term.

\section{Stability of Runge-Kutta methods}
\label{sec:rkmethods}

The MFMS model Eq.~\eqref{eq:system} has been explicitly formulated as a sum of three types of terms: diffusion terms ($\Fdiff$),  advection terms ($\Fadv$), and stiff reactive terms ($\Freac$). Both types of terms $\Fdiff$ and $\Fadv$ display spatial connectivity, whereas the stiff source terms ($\Freac$) lack spatial connectivity and feature timescales considerably smaller than those associated with other terms.

In this section, following \citet[Section IV]{Hairer1996}, we concentrate on the Dalquist test problem to discuss the time integration of these three types of terms. We review temporal integration methods, predominantly Runge-Kutta methods, emphasizing current effective and accurate approaches.

\subsection{The Dalquist equation}

Consider the so-called Dahlquist test equation
\begin{equation} \label{eq:Dalquist}
\frac{dy(t)}{dt} = F(y(t))=\lambda y(t),\ y(0)=1,
\end{equation}
where $\lambda \in \mathbb{C}$ is a fixed parameter corresponding to an eigenvalue of the system obtained after linearizing a nonlinear stiff system of differential equations. It is well understood that in the considered  Eq.~\eqref{eq:system}, the function $F$ from Eq.~\eqref{eq:Dalquist} plays the role of the respective terms $\Fadv$, $\Freac$, or $\Fdiff$.  In this context, for these three cases, the eigenvalue $\lambda$ will exhibit different characteristics, such as being purely imaginary ($\Fadv$), 
complex with a dominant negative real part ($\Fdiff$), or purely real ($\Freac$).

Applying to Eq.~\eqref{eq:Dalquist} a Runge-Kutta method with timestep $\Delta t$ yields a recursion of the form
\begin{equation} \label{eq:defR}
y_{n+1} = R(z) y_n,
\end{equation}
where $z = \lambda \Delta t$ and $R(z)$ is a rational function called the stability function of the Runge-Kutta method. The numerical solution $y_n$ in Eq.~\eqref{eq:defR} remains bounded for $n\rightarrow +\infty$ if and only if
$z\in S$, where we defined the stability domain of the Runge-Kutta method as
\begin{equation} \label{eq:defS}
S=\{z\in \mathbb{C}\ ;\ |R(z)| \leq 1\}.
\end{equation}
Depending on the nature of an evolution problem ($\Fadv$, $\Freac$, or $\Fdiff$), different Runge-Kutta methods can be more suitable. We will review them in the following subsections.

\subsection{Explicit stabilized Runge-Kutta methods for diffusion problems}
\label{sec:diffusion}

Let us focus first on diffusive terms applied to the Dahlquist test in Eq.~\eqref{eq:Dalquist} with $F=\Fdiff$. The eigenvalue $\lambda$ is close to the negative real axis in the complex plane. Tackling the integration of diffusion operators presents challenges due to the numerical stiffness induced by the CFL condition stemming from the involvement of second-order differential operators. 

For the explicit Euler method \( y_{n+1} = y_n + \Delta t F(y_n) \), the stability function \( R(z) = 1 + z \) defines a stability domain \( S \) as a disk centered at \(-1\) with radius 1, encompassing the real interval \([-2, 0]\). This results in a stability condition \( \Delta t \leq 2/|\lambda| \), where \( \lambda \) denotes the largest eigenvalue. When diffusive terms are discretized using standard finite difference methods with a spatial mesh size \( \Delta x \), the largest eigenvalue typically scales as \( |\lambda| \leq C\Delta x^{-2} \). This imposes a severe CFL condition to ensure numerical stability.

A natural approach to circumvent this timestep restriction and stability for diffusion problems is to use implicit methods, but these can suffer from a prohibitive cost and difficulties in implementation due to the need for sophisticated linear algebra solvers with preconditioners in Newton-type nonlinear solvers \citep{Hairer1996}.

\begin{figure*}[!tbh]
	\begin{center}
	   \hspace*{-0.15cm}
		\includegraphics[width=0.9\linewidth]{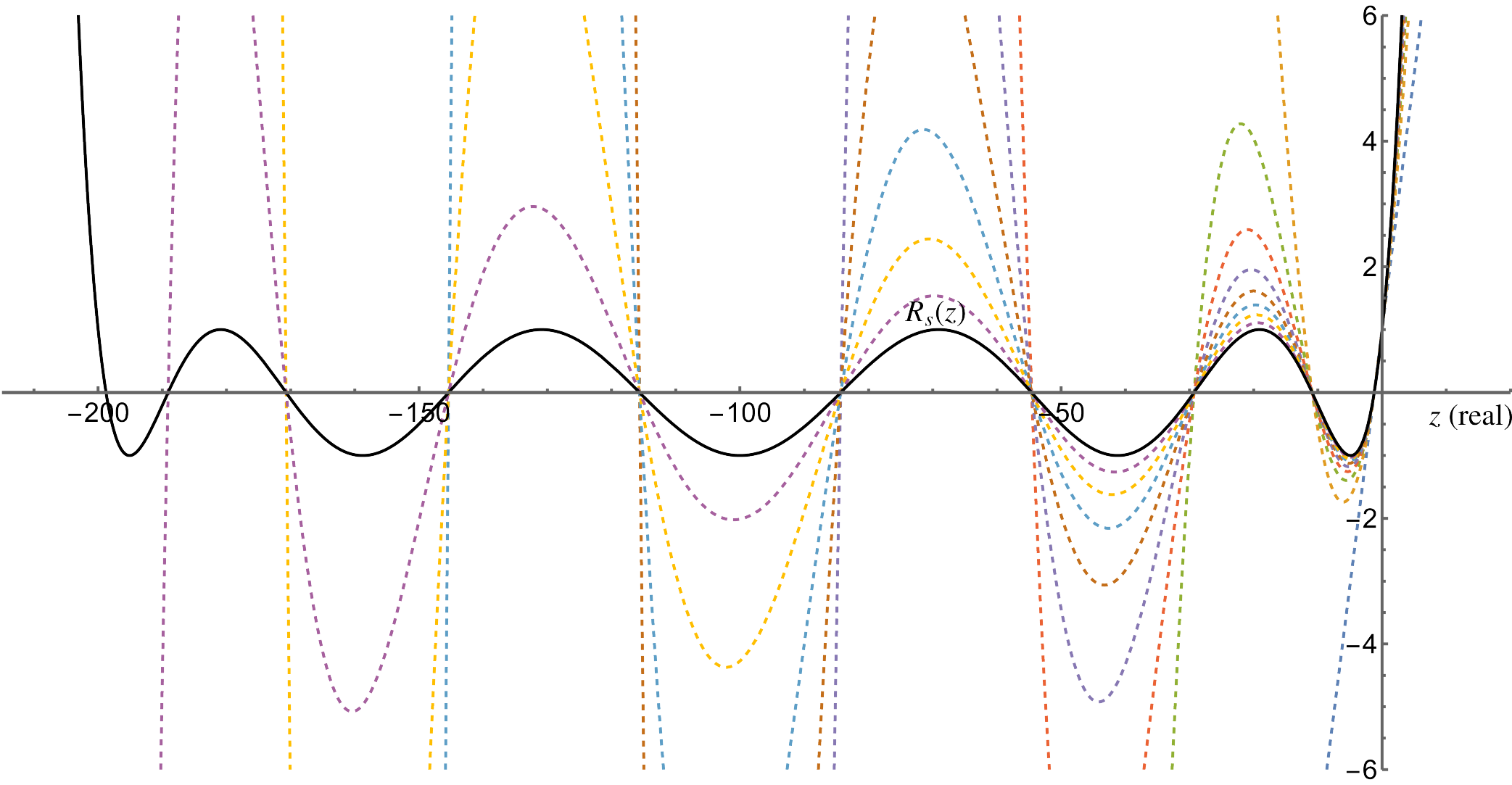}\\[5mm]
  \includegraphics[width=0.9\linewidth]{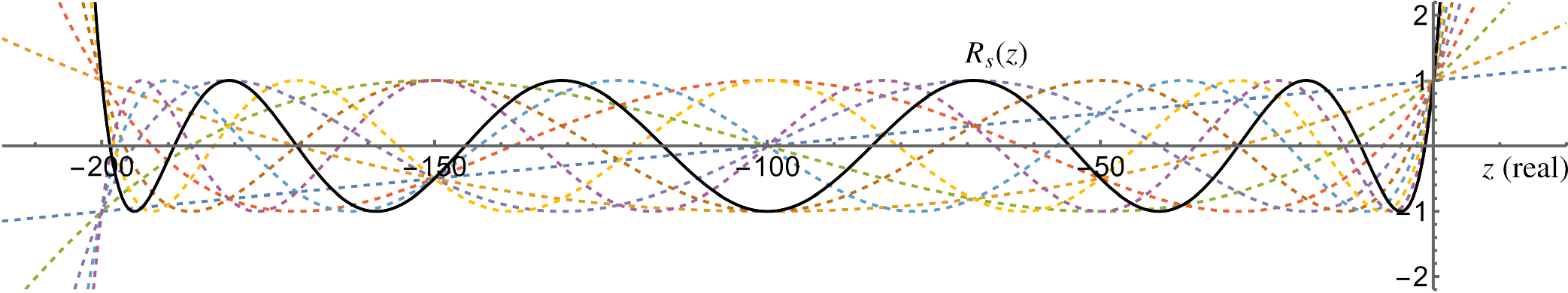}
		\caption{Comparison of the STS implementation in Eq.~(\ref{eq:defnaiv}) (top) and the stable implementation based on Chebyshev recurrence in Eq.~(\ref{eq:defrec}) (bottom) of the first-order Chebyshev method with optimal stability polynomial given in Eq.~(\ref{eq:defRs}).
  The stability polynomial $R_s(z)$ defined in Eq.~(\ref{eq:defRs}) for $s=10$ (see black curves) and the stability polynomials of internal stages (related to $k_j,j=1\ldots,s-1$, see color curves) are plotted as a function of $z$ which is assumed purely real here.
  The STS implementation exhibits poor internal stability, with internal stability polynomials oscillating with large amplitude resulting in instabilities with respect to roundoff errors, in contrast to the Chebyshev recurrence implementation defined in Eq.~\eqref{eq:defrec}, with favorable internal stability where the amplitude of the internal stage stability polynomials remains bounded by $1$.}
		\label{fig:stabilityRKC}
	\end{center}
\end{figure*}

\subsubsection{Limitations of the Super Time Stepping Method}

As an alternative to implicit schemes, the use of explicit stabilized methods can be very efficient for diffusion problems, as presented in \citet[Section IV.2]{Hairer1996} (see also the review by \citet{Abdrev}). The idea is to consider explicit schemes with stability function defined in Eq.~\eqref{eq:defR} that have extended stability domains $S$ along the negative real axis as defined in Eq.~\eqref{eq:defS}. 
To achieve such improved stability properties, the main idea is to allow the variation of the number $s$ of the internal stage of the methods that then grow adaptively with the required stability. This is in contrast with standard Runge-Kutta methods, where the number of internal stages is, in general, constant and has to be increased to improve the order of accuracy.
For a fixed $s\geq 1$, among all polynomials $R_s(z)=1+z+a_2z^2+a_3z^3+\ldots +a_sz^s$ of degree at most $s$, corresponding to consistent explicit $s$-stage Runge-Kutta methods, the polynomial that maximizes the length $L_s$ of the stability domain $[-L_s,0]\subset S$ is given by 
\begin{equation} \label{eq:defRs}
R_s(z)=T_s(1+\frac{z}{s^2})
\end{equation}
and benefits from the large stability domain size $L_s=2s^2$ along the negative real axis and $T_s$ denotes the Chebyshev polynomial of degree $s$ (for instance $T_0(x)=1, T_1(x)=x,T_2(x)=2x^2 -1$, \ldots). This is shown using the property $T_s(\cos \theta)=\cos(s\theta)$ for all $\theta$ that characterizes Chebyshev polynomials. This method relaxes the CFL condition of diffusion problems and allows for larger timesteps. 
It entails an increase in internal stages, resulting in a quadratically expanding the size $L_s$ of the stability domain along the negative real axis, rendering it more stable than standard explicit methods, computationally cheaper and easier to implement than implicit methods. 
Precisely, the stability condition becomes for diffusion problems
$
\Delta t |\lambda| \leq L_s=2s^2
$
which yields the relaxed CFL condition $\Delta t \leq Cs^2 \Delta x^2$. 

Then, a naive implementation is to implement such explicit stabilized methods as a composition of explicit Euler steps, as proposed independently by \cite{Saulev60} and \cite{Guillou60},
\begin{equation} \label{eq:defnaiv}
k_j = k_{j-1} + \delta_j \Delta t F(k_{j-1}),\quad j=1,\ldots, s,
\end{equation}
with $k_0=y_n$, and the output $y_{n+1}=k_s$ coincides with the last internal stage, corresponding to the factorization
$
R_s(z) = \prod_{j=1}^s (1+\delta _j z),
$
where $-1/\delta_j,j=1\ldots,s$ denote the $s$ roots of the optimal stability polynomial $R_s$ defined in Eq. \ref{eq:defRs}  \citep[i.e. $R_s(-1/\delta_j)=0$ for all $j=1\ldots,s$, see again the review by][]{Abdrev}. While the ordering of the roots can be optimized, this naive approach still results in first-order methods with unstable internal stages. As a result, these methods are typically practical only for small degrees 
$s$, generally less than ten. This approach was rediscovered as the Super Time Stepping (STS) method \citep[see][]{Genevi96super-time-steppingacceleration} and commonly used in the context of the temporal integration of the ambipolar diffusion in single-fluid MHD models by \citep[e.g.][]{Genevi96super-time-steppingacceleration,Gonzalez_2018,Nobreg-Siverio:2020AA...633A..66N,Mignone2007ApJS..170..228M}.

\subsubsection{Second-order orthogonal Chebyshev method}

The stability issues related to round-off errors, primarily due to instabilities in the internal stages, were effectively addressed in the seminal work by \cite{vanderHouwen1980}. They resolved this by computing the internal steps of the Runge-Kutta methods using a recurrence formula analogous to the recurrence formula $T_{j}(x)=2xT_{j-1}(x)-T_{j-2}(x)$ for all $j\geq 2$ of Chebyshev polynomials. The internal steps are then given by:
\begin{equation}
k_j = \frac{2\Delta t}{s^2} F(k_{j-1}) + 2k_{j-1} - k_{j-2}, \quad j=2,\ldots, s,
\label{eq:defrec}
\end{equation}
where \( k_0 = y_n \), \( k_1 = y_n + \frac{\Delta t}{s^2} F(y_n) \), and the output \( y_{n+1} = k_s \) coincides with the last internal stage.

Although both implementations in Eq. \eqref{eq:defnaiv} and Eq. \eqref{eq:defrec} share the same stability function (Eq. \eqref{eq:defRs}), making them equivalent methods for a linear vector field $F$ and in the absence of roundoff errors, the advantage of the implementation in Eq.~\eqref{eq:defrec} based on the Chebyshev recurrence compared with the STS method in Eq.~\eqref{eq:defnaiv} is illustrated in Fig.~\ref{fig:stabilityRKC}.  Here, favorable internal stability is observed compared to the naive implementation in Eq.~\eqref{eq:defnaiv}, where the stability polynomials of the internal stages exhibit large amplitude oscillations. This issue is evident for \(s=10\) and worsens as the number of internal stages \(s\) increases, making Eq. \eqref{eq:defnaiv} unusable for larger numbers $s$ of stages due to its unstable behavior already with respect to roundoff errors, as analyzed by \cite{vanderHouwen1980}.

In particular, the main contribution in \cite{vanderHouwen1980} is a class of second-order explicit stabilized methods based on Chebyshev polynomials with favorable internal stability and error control, known as the RKC method. This work inspired new classes of explicit stabilized methods with quasi-optimally large stability domains, notably the explicit second-order orthogonal Chebyshev method known as ROCK2 based on appropriate families of orthogonal polynomials\footnote{Publicly available at \cite{ABDULLE2013869CODE}} \citep{abdulle_2008, Abdulle2002, ROCK2_2001, Zbinden2011}.

The ROCK2 method shares a concept similar to the STS method, featuring a quadratically increasing stability domain, but boasts three significant advantages. Firstly, STS is only first order, while ROCK2 is a second-order method with a stability condition \(\Delta t |\lambda| \leq C s^2\), where \(C=0.81\) is quasi-optimal among second-order explicit stabilized methods (see the stability domain as defined in Eq. \eqref{eq:defS} of length $L_{13} \simeq 136$ in the left picture of Fig. \ref{fig:stabilityPIROCK}). Secondly, ROCK2 can handle large internal stages (up to 200 is implemented) without encountering stability issues, thanks to a recurrence based on orthogonal polynomials, compared to the ten-stage limit for STS. Lastly, ROCK2 includes an embedded method, enabling seamless integration with an error estimator for optimal timestep estimation, as detailed in Section~\ref{sec:timestep}. Note that the hyperdiffusive term can be implemented and benefit from this method. Notably, it does not require a precise spectral radius estimation but only an upper bound, providing a significant computational cost advantage. Further details can be found in Appendix~\ref{app:rock2}.

\begin{figure*}[!tbh]
	\begin{center}
	   \hspace*{-0.4cm}
		\includegraphics[height=3.85cm]{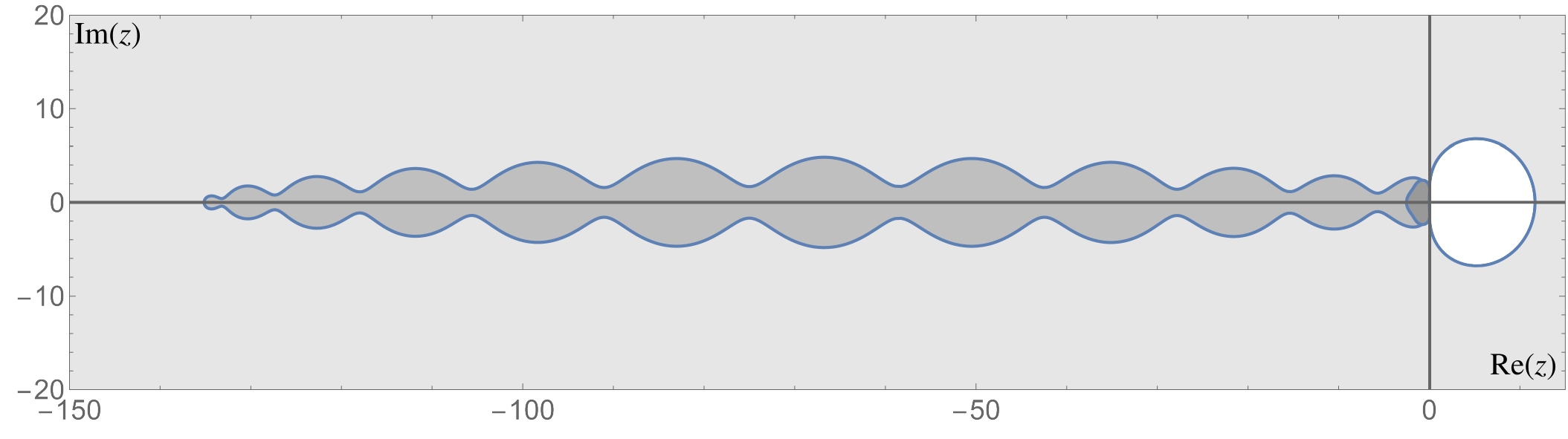}
  \hspace{1mm}
  \includegraphics[height=3.9cm]{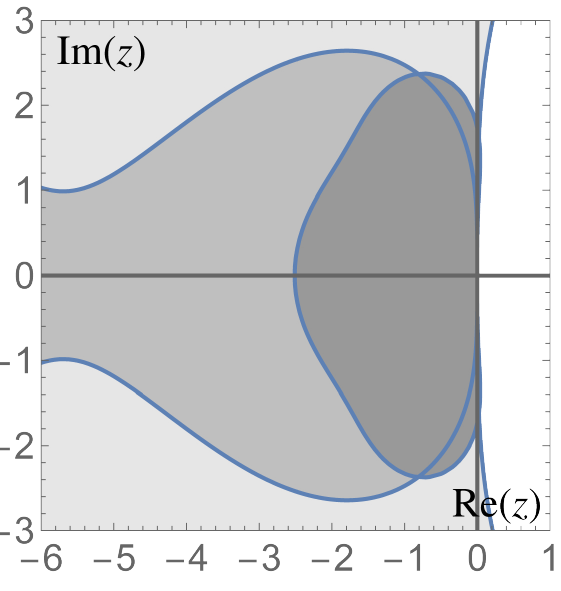}
		\vspace*{0.3cm}
	\caption{Stability domains as defined in Eq.~\ref{eq:defS} of the schemes involved in PIROCK: ROCK2 (in gray. here $s=13$ internal stages) for diffusion terms, an explicit method of order $3$ for advection terms (dark gray), and the L-stable diagonally implicit Runge-Kutta method of order $2$ for stiff reaction terms (light gray). Right picture: zoom near the origin to illustrate that the advection method (dark gray) includes the portion of the imaginary axis. The stability domains are plotted for $z$ in the complex plane (horizontal axis: real part $\mathrm{Re}(z)$, vertical axis: imaginary part $\mathrm{Im}(z)$).}
		\label{fig:stabilityPIROCK}
	\end{center}
\end{figure*}

\subsection{Runge-Kutta methods for advection problems and Diagonally implicit Runge-Kutta methods for stiff reaction problems}
\label{sec:sourceadv}

We focus on the Dalquist test for both advection ($\Fadv$ in Eq.~\eqref{eq:system}) and stiff reaction problems ($\Freac$), as described by Eq.~\eqref{eq:Dalquist}.
We recall that the constant $\lambda$ in Eq.~\eqref{eq:Dalquist} plays the role of an eigenvalue of a linearized version of the vector field,
 which is purely imaginary for advection problems and real for stiff reaction problems. 

As explained in the introduction and considered by many other codes, classical Runge-Kutta methods have demonstrated effectiveness and robustness for pure advection problems. We opt for a classical third-order explicit Runge-Kutta Heun method in our proposed scheme for advection problems. Indeed, it turns out that the stability domain of explicit Runge-Kutta methods with an odd order of accuracy includes a portion of the imaginary axis in a neighborhood of the origin. This method is characterized by embedded error estimation, which allows for assessing solution errors without requiring extra evaluations. This capability is attained through a second-order approach that combines the internal stages utilized in the third-order method implemented in our approach.

For stiff reaction problems, source terms are often integrated implicitly due to the absence of spatial connectivity and high numerical stiffness. Indeed, note that Ebysus, like Bifrost, uses a staggered mesh, therefore, all the terms of $\Freac$ are centered a priori. 
Our approach employs a two-stage second-order Singly Diagonally Implicit Runge-Kutta (SDIRK) method, known for its L-stability and efficient implementation \citep[see][]{Alexander1977,Hairer1996}. L-stability means that the stability domain defined in Eq. \eqref{eq:defS} contains the left half place $\{\Re (z) \leq 0\}$ and in addition $R(\infty)=0$ which is a desirable property for severely stiff problems.
The diagonal structure of the matrix of Runge-Kutta coefficients allows for a single lower-upper (LU) factorization per time step when using a quasi-Newton method, as elucidated by \citet{Alexander1977} and \citet{Hairer1996}. The integration method is detailed in Appendix~\ref{app:DIRK}. Notably, given the nonlinearity of the source terms in Eq.~\eqref{eq:system}, the Jacobian has been calculated numerically using finite differences to avoid computing derivatives. Note that such a numerical Jacobian approximation does not affect the accuracy of the implicit Runge-Kutta method used for the reaction terms, but only possibly its nonlinear implementation using a Newton method (\citet[Section IV.8]{Hairer1996}).

\subsection{Time integration strategies for stiff diffusion-reaction-advection problems}

In the preceding subsections, we reviewed the methods we employed to individually integrate each term, namely $\Fadv$, $\Freac$, and $\Fdiff$, within the framework of the Dalquist problem. However, a significant challenge lies in devising a temporal integration method that effectively handles all three terms simultaneously. Previously, we discussed various strategies for handling all three terms simultaneously. In particular, we mentioned the time operator splitting technique, as considered in \citet{strang,duarte,theseduarte,Splitting}. 

An emerging strategy employs the partitioned implicit-explicit RK methods (IRKC) method, as described by \citet{Verwer2004}. Additionally, there is a modified version, as described by \citet{SOMMEIJER20073}, as a two-step method with enhanced imaginary stability, and also a second-order partitioned method PRKC \citet{Zbinden2011}.

In this study, we consider a novel partitioned implicit–explicit integrator named PIROCK, as described by \citet{ABDULLE2013869}. PIROCK combines each time integration method for individual terms outlined in the previous subsections~\ref{sec:diffusion} and \ref{sec:sourceadv}. The selection of PIROCK is driven by its versatility and efficiency, serving as a ``swiss-knife" solution for handling various regimes of the typical problem described in Eq.~\eqref{eq:system} with a single code. In the context of integrating the MFMS system, the PIROCK approach emerges as particularly apt. Diverging from conventional splitting methods, the PIROCK approach incorporates a timestep control strategy based on error estimators, thereby facilitating the optimization of the timestep while guaranteeing stability. Consequently, we circumvent the inherent errors associated with splitting methods.

\section{The PIROCK method}~\label{sec:numerical}

In this section, we provide a concise overview of the PIROCK algorithm proposed by \citet{ABDULLE2013869}. 
 
 \subsection{Overview of PIROCK}

PIROCK\footnote{The source code of PIROCK is publicly available at \cite{ABDULLE2013869CODE}} is a unified second-order partitioned Runge-Kutta method that amalgamates the RK integration methods ROCK2, Heun, and SDIRK discussed in Sections~\ref{sec:sourceadv}, and \ref{sec:diffusion}, respectively. The construction details of PIROCK are described in-depth in the work by \citet{ABDULLE2013869}. A description of the algorithm is given in Appendix~\ref{app:pirock}. 
 
 The PIROCK method performs the diffusion step first and introduces the advection and reaction steps as a finishing procedure, as described in Appendix~\ref{app:pirock}. This diffusion step is coupled with the $\Fadv$ and $\Freac$ methods, following a choice of coupling that is not unique. This particular coupling has been selected because it has the same computational cost as three non-partitioned RK methods, corresponding to the presented ROCK2, Heun, and SDIRK methods but with two additional evaluations of the diffusion function $\Fdiff$ required at each step. The PIROCK method necessitates $s+3$ evaluations of $\Fdiff$, where $s$ is the number of internal stages of the ROCK2 method, along with three evaluations of $\Fadv$ and two resolutions of a nonlinear system of $\Freac$ at each time step.

When $s=1$, indicating a scenario where diffusive processes have limited influence, or when the spectral radius of the Jacobian of $\Fdiff$ is small compared to the spectral radius of the Jacobian of $\Fadv$, the PIROCK method simplifies to a standard non-partitioned third-order RK scheme. In this context, advective and diffusive terms are explicitly evaluated using the Heun method, while source terms are implicitly integrated with the SDIRK approach (Appendix~\ref{app:pirock}).

It is noteworthy that Shampine (see \citet[Section~IV.8]{Hairer1996}) previously proposed the utilization of multiplying certain steps by $\bold J^{-1}_{\text{R}}$, where $\bold J^{-1}_{\text{R}}$ denotes the inverse of the Jacobian matrix associated with the source terms $\Freac$, as a means of stabilization. In this approach used in PIROCK, the factor $\bold J^{-1}_{\text{R}}$ is also used to stabilize further the coupling of the diffusion step with the $\Fadv$ and $\Freac$ methods as described by \citet{ABDULLE2013869CODE}.

 \subsection{Timestep calculation, embedded methods, and number of stages of the ROCK2 method}
 \label{sec:timestep}

This section discusses the procedure for calculating the timestep and the number of stages of the ROCK2 method. The chosen approach is based on an adaptive timestep that takes into account the errors using the idea of embedded methods introduced by \citet{Hairer1993}.

Runge-Kutta methods, including the PIROCK method, can be characterized by coefficients that define the internal stages, as presented in the Butcher tableau \cite[see][]{BUTCHER1996247}. To harness the benefits of embedded methods, a novel set of coefficients is introduced as a perturbation to this Butcher tableau. These newly defined coefficients are chosen to deliberately violate the second-order non-partitioned conditions while adhering only to the first-order conditions—except for advection methods, which incorporate a second-order condition in the PIROCK method. The determination of these coefficients relies on algebraic constraints derived exclusively from these conditions. Finally, the error estimators are computed straightforwardly by comparing the first and second-order approximations of each integrator, incurring no additional costs. The error is a combination of the already calculated internal stages of the RK method.

During each time integration step, if $s>1$, the error estimators are computed as a combination of the internal stages of the PIROCK method described in Appendix~\ref{app:pirock}, covering the entire spatial domain, as follows:

\begin{equation}
\begin{aligned}
	\text{err}_D &=\Delta t\,\sigma_{s} (1-\tfrac{\tau_{s}}{\sigma^2_{s}})\lVert\Fdiff(\widetilde{\consvar}_{s-1}) - \Fdiff\left(\consvar_{s-2}\right)\rVert,  \\ 
    \text{err}_A &=\tfrac{\Delta t}{10}\lVert-\tfrac{3}{2}\Fadv(\consvar_{s+1})+3\Fadv(\consvar_{s+4})-\tfrac{3}{2}\Fadv(\consvar_{s+5})\rVert^{\tfrac{2}{3}},\\
    \text{err}_R &= \tfrac{\Delta t}{6}\lVert\bold J^{-1}_{\text{R}}\left(\Freac(\consvar_{s+1})-\Freac(\consvar_{s+2})\right)\rVert.
		\end{aligned}
\label{eq:errorest}
\end{equation}
where $\sigma_{s}$ and $\tau_{s}$ are coefficients derived from the Chebyshev recurrence relation of orthogonal Chebyshev polynomials, calculated to achieve second-order accuracy for any given total number of stages $s$,  $\lVert . \rVert$ denotes the L2 norm, 
the power $2/3$ in the definition of $\text{err}_A$ accounts for the order 3 of the considered explicit method for advection terms compared to the order 2 of PIROCK, and $\widetilde{\consvar}_{s-1}$ corresponds to the solution of $\consvar$ during the finishing procedure for diffusion (as shown in Appendix \ref{app:pirock}).

Subsequent to this, the global error is straightforwardly defined as:
 \begin{equation}
     \text{err} = \max (\text{err}_\text{D}, \text{err}_{\text{A}}, \text{err}_{\text{R}}).
\end{equation}

The error estimators facilitate adaptive timestep adjustments to maintain the solution error within a user-prescribed tolerance $tol$, aiming for $\text{err} \simeq tol$. Two scenarios arise in this context. If the error exceeds the tolerance ($\text{err}>\text{tol}$), the proposed test is rejected, prompting consideration of a new smaller timestep to achieve $\text{err} \simeq tol$. Conversely, when the error is well below the tolerance, an increase in timestep is warranted. Estimating the timestep accounts for these scenarios to achieve the prescribed error tolerance $tol$ while avoiding too many rejected steps. 

The timestep estimation is refined slightly in \citet{ABDULLE2013869} by considering the previous timestep $\Delta t^{p}$. This modification is standard for stiff problems \citep[Section~IV.8]{Hairer1996} to avoid undesired oscillations in the timestep sizes across steps, resulting in a smooth evolution of the timestep. In this context, the next timestep $\Delta t^n$ is predicted as 

 \begin{equation} \label{eq:timestepchoice}
     \Delta t^n= 0.8\, \Delta t \, \sqrt{\frac{\text{tol}}{\text{err}}}\min\left(1,\,\frac{\Delta t}{\Delta t^p}\sqrt{\frac{\text{err}^p}{\text{err}}}\right),
 \end{equation}
where $\text{err}^p$ is the error calculated from the previous step and $0.8$ is a standard safety factor for stiff problems. Note that if $\text{err}$ is very small compared to the tolerance, there might be a temptation to choose a large timestep. However, it is crucial to remember that in addition to the timestep selection strategy in Eq.~\eqref{eq:timestepchoice} the maximum timestep in PIROCK is determined by the spectral radius of the Jacobian of the advective terms $\Fadv$. The error estimator may underestimate errors when CFL conditions imposed by the advective terms are violated, so we ensure the CFL constraint is always respected. 

Ultimately, the PIROCK method's number of stages ($s$) is approximated based on the spectral radius estimation of the diffusive term, denoted as $R_\text{D}$, using:
\begin{equation} \label{eq:defs0.43}
s \simeq \sqrt{\frac{\Delta t \, R_D}{0.43}},
\end{equation}
where the estimation of the spectral radius for the diffusive term $\Fdiff$ in the system of equations in Eq.~\eqref{eq:system} follows a conventional power method, commonly referred to as the Von Mises iteration \citet{MisesPraktischeVD}.

\section{Results}~\label{sec:tests}

In this section, we comprehensively assess the PIROCK method as applied to the system of equations~\eqref{eq:system} by examining 1D and 2.5D scenarios. We systematically evaluate the performance of the PIROCK method with tolerances from $tol=10^{-2}$ to $tol=10^{-5}$ in comparison to two established numerical approaches:

\begin{itemize}
    \item The ``explicit" approach, denoted as explicit, employs a full explicit integration strategy. In this approach, all terms of Eq.~\eqref{eq:system} are integrated explicitly, including the diffusive ($\Fdiff$) and source ($\Freac$) terms, employing an RK3-2N method.
    \item The ``operator splitting" approach, denoted as splitting, follows the methodology as outlined by \citet{wargnier2022multifluid}. In this approach, $\Fadv$ and artificial hyperdiffusive terms are integrated using an RK3-2N method, the Spitzer term is integrated using a wave method, as described by \citet{Rempel_2016}, and the source terms are integrated implicitly through a fifth-order implicit Runge Kutta method also known as \textit{Radau IIA} method \citep[see][]{HAIRER199993,Hairer1996}. The coupling of these terms is handled using a first-order Lie splitting approach.
\end{itemize}

The following comparative analysis demonstrates the suitability of the PIROCK method for both multi-fluid and single-fluid MHD models and its potential benefits in terms of computational cost savings.

We initially focus on a 1D single-fluid Sod-Shock tube problem for this comparison. We compare the explicit approach with PIROCK for different tolerances, evaluating the computational cost versus accuracy between the two methods. We aim to comprehend how PIROCK determines the number of stages $s$ for this type of problem and how it stabilizes the scheme while significantly reducing computational cost by executing fewer steps than the explicit approach.

In the following investigation, we explore a 2.5D reconnection scenario utilizing an MFMS model. Our analysis begins by examining various species involved, including $\{\text{H}, \text{H}^+, \text{He}, \text{He}^+ \elec\}$. Through comparative analysis, we assess the performance of the PIROCK method with different tolerances and compare it with the splitting and explicit.

\subsection{Sod-Shock}

In this section, we analyze a Sod-Shock problem following \citet{Sod:1978fj}. We compare PIROCK cases (with tolerances from $tol=10^{-2}$ to $tol=10^{-5}$ and an extra case at $tol=10^{-8}$) with an explicit calculation to evaluate their performance in a purely hyperbolic problem. We consider only the advective part of the system of Eqs.~\eqref{eq:system} for a 1D hydrodynamic single-fluid with artificial hyperdiffusion. Therefore, the system involves a single set of continuity, momentum, and energy equations, represented by the conservative variables vector $\consvar = \left(\rho, \rho u_x, e\right)$. 

\subsubsection{Initial conditions}

The initial left and right state of the vector $\consvar$ in the Sod-Shock problem is described by 

\begin{equation}
\consvar_{L}=\left(1,\, 0,\,\tfrac{1}{(\gamma - 1)}\right)^T \, \text{and} \, \consvar_{R}=\left(0.125,\, 0,\,\tfrac{0.1}{(\gamma - 1)}\right)^T
\end{equation}
where $\gamma = 1.4$. We consider a smooth initial solution where the two states $\consvar_{L}$ and $\consvar_{R}$ are connected by the following hyperbolic tangent 

\begin{equation}
    \bold f (x) = \left(\tfrac{\consvar_{L}+\consvar_{R}}{2}\right)+\left(\tfrac{\consvar_{L}-\consvar_{R}}{2}\right)\tanh{\left(\tfrac{x - 0.5}{\lambda}\right)},
\end{equation} 
where $x \in [0, 1] $ and $\lambda = 10^{-2}$. We ran simulations for all PIROCK and explicit cases until $t=t_f=0.2$. We consider a uniform mesh of 256 grid points. 

A reference solution was constructed utilizing the explicit case with an extremely small fixed timestep of $\Delta t = 10^{-9}$ for method comparison. Additionally, we compared our results with the classical analytical solution of the Sod-Shock problem that considers discontinuous initial conditions instead of a smooth hyperbolic tangent.

To be consistent with what is generally used in \bifrost\, \citep[e.g.,][]{Martinez_Sykora_2020,Nobrega-Siverio:2016qf,Nobrega-Siverio:2017sim}, we have chosen the hyperdiffusive terms coefficients $\nu_1 = \nu_2 = 0.2$ and $\nu_3 = 0.3$ in Eq.~\eqref{eq:hypdiff} for all cases.

\subsubsection{Comparative analysis of all methods}

Figure~\ref{fig:1dsodshock} illustrates the density distribution for all cases. The top-left panel displays the distribution of numerical solutions for the entire domain, while the top-middle panel shows the difference from the reference solution. The top-right panel shows the difference between the reference and analytical solution of the density. The bottom panels zoom on the region between $x=0.82$ and $0.88$, indicated by the two vertical blue lines around the shock wave in the top-left panel. This emphasis highlights differences in the solution in this region, where the error and $L_2$ norm are maximized.

All cases (including various tolerances of PIROCK and the explicit case) successfully capture the characteristic waves and jump conditions of the Sod-Shock test case, as compared with the analytical solution (top-left panel of Fig.~\ref{fig:1dsodshock}). The tolerance decreases, and the difference between the numerical solution for the PIROCK cases and the reference solution diminishes (top-middle panel). This trend is further emphasized when focusing on the shock region (bottom row of Fig.~\ref{fig:1dsodshock}). Higher tolerances ($10^{-2}$ and $10^{-3}$) yield a slightly more diffusive solution compared to lower tolerances.

Table~\ref{tab:sodshock} compares various parameters between the explicit and PIROCK cases for different tolerances. Specifically, we assess the number of steps to reach $t=t_f$, the average timestep $\overline{\Delta t}$ chosen by each method, the number of evaluations for the hyperdiffusion term $\Fdiff$ and the advection term $\Fadv$, the maximum number of stages $s$ selected by PIROCK during the simulation, the L2 norm between the numerical and reference solutions for the density variable $\rho$, the average CPU time per step for each case and the total computational time.

In our study, we observed that the explicit case necessitates 2203 steps to achieve an average timestep of $\Delta t =8.76\times 10^{-5}$, yielding a $L_2$ norm of the density of $2\times 10^{-6}$, with a total computational time of 13.2 s. Transitioning to the PIROCK case with the highest tolerance ($10^{-2}$ or $10^{-3}$), we managed to reduce the total computational cost to 3.8 s, requiring only 160 steps. However, this optimization comes at the expense of a higher $L_2$ norm of the density, measured at $7.6\times 10^{-5}$. As the tolerance decreases further ($10^{-4}$ or $10^{-5}$), we noted an increase in computational cost (from 3.9 to 5.1 s) alongside a decrease in the $L_2$ norm (ranging from $3.4\times 10^{-5}$ to $10^{-5}$). This trend aligns with the underlying principles of the PIROCK approach, which is expected to refine the solution as the tolerance diminishes. Additionally, our findings suggest the possibility of achieving a superior approximation of the solution compared to the explicit 3rd order method by further reducing the tolerance to $10^{-8}$.

As depicted in Table~\ref{tab:sodshock}, an increase in tolerance prompts a corresponding increase in the maximum number of stages ($s_{max}$). Specifically, for tolerances ranging from $10^{-4}$ to $10^{-2}$, the number of stages rises to 10 and 12, respectively, compared to mere counts of 1 and 4 for tolerances of $10^{-8}$ and $10^{-5}$. This observed behavior is attributed to PIROCK's dynamic adjustment of the stage number $s$ while maintaining an approximation error ($err$) close to the prescribed user tolerance ($tol$). When the tolerance is increased, the PIROCK algorithm allows for a larger timestep and optimally adjusts the number of stages while remaining in the stability domain of the diffusive terms. Note that the explicit method is restricted not only from the CFL condition of the advective term but also from the hyperdiffusive terms. However, PIROCK solves the hyperdiffusive terms with the ROCK2 method.

In summary, PIROCK allows for fine-tuning solution quality and computational costs through user-defined tolerance. The method demonstrates superior efficiency, requiring substantially fewer steps than the explicit approach. The explicit approach takes 13.3 s compared to 3.8 seconds for the highest tolerance $tol=10^{-2}$. The achieved solution quality remains highly satisfactory relative to the number of steps taken, establishing PIROCK as a more efficient alternative to the explicit approach even for a classical Sod-Shock test.

\begin{figure*}[!tbh]
	\begin{center}
	   \hspace*{-0.4cm}
		\includegraphics[scale=0.35]{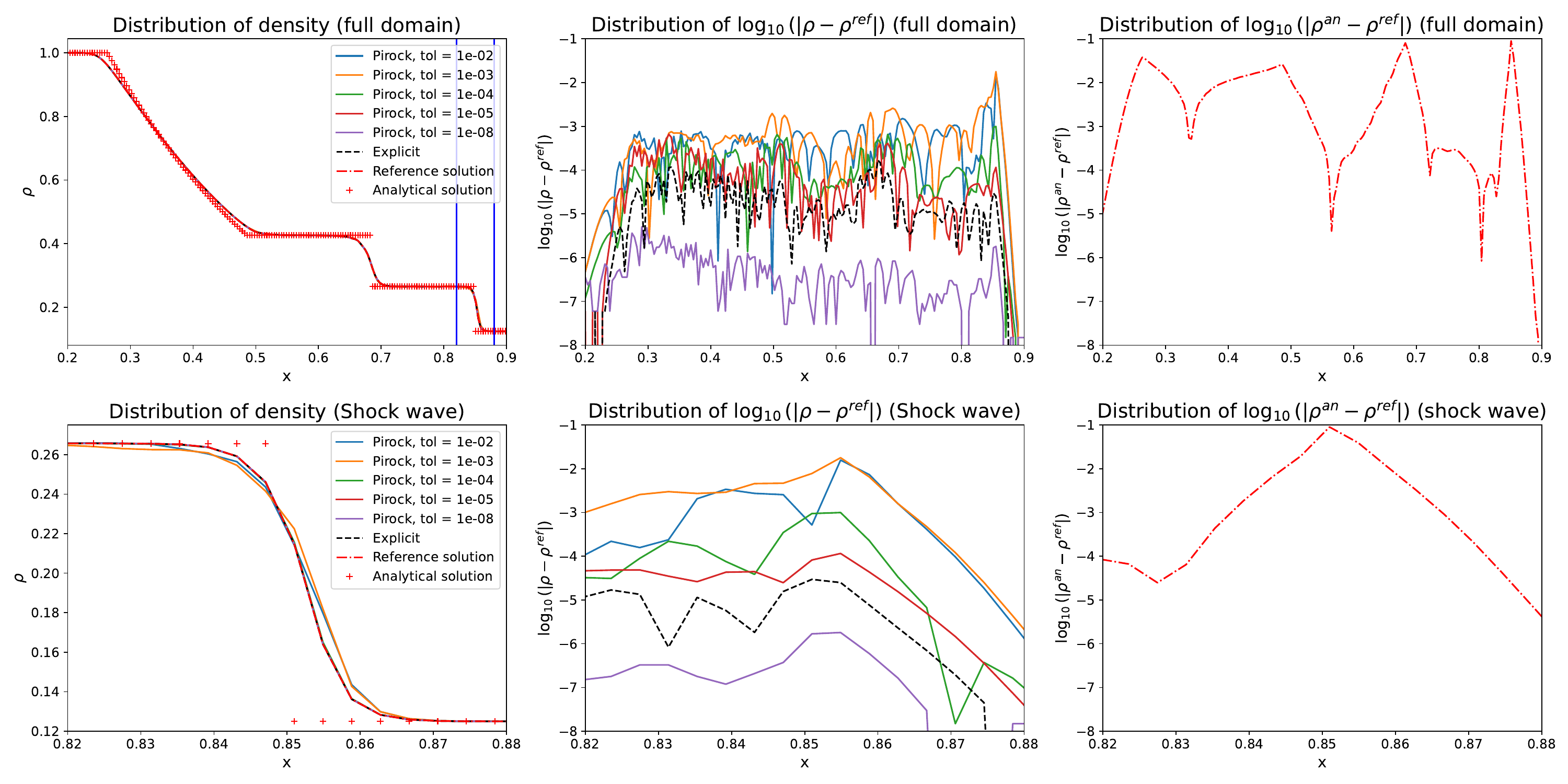}
		\vspace*{0.1cm}
		\caption{Density distribution across the entire domain for PIROCK with different tolerances, the explicit approach, and reference solutions (top left). The top-middle panel depicts the distribution of the logarithm of the absolute difference, $\log_{10}(|\rho-\rho^{ref}|)$, between the different scenarios and the reference solution. The top-right panel shows the distribution of the logarithm of the absolute difference between analytical and reference solution $\log_{10}(|\rho-\rho^{ref}|)$. The bottom panels mirror the top ones but focus specifically on the shock region between x=0.82 and 0.88 (indicated by the two vertical blue lines in the top-left panel). The red dashed line represents the reference solution, and the black dashed line corresponds to the explicit. The remaining colors represent different PIROCK cases with varying tolerances.}
		\label{fig:1dsodshock}
	\end{center}
\end{figure*}

\begin{table*}
\begin{tabular}{c|c|c|c|c|c|c|c|c|}
\cline{2-9}
& Steps (rej. steps) & $\overline{\Delta t}$& $\Fdiff$ evals & $\Fadv$ evals & $s_{\text{max}}$ & $L_2$ norm            & CPU time/step [s] & Tot. time [s] \\ \hline
\multicolumn{1}{|c|}{explicit}                 & 2203 (0)                     & $8.8\times 10^{-5}$                             & 6609       & 6609       & 0          & $2\times 10^{-6}$ & $6\times 10^{-3}$ & 13.3\\ \hline
\multicolumn{1}{|c|}{PIR. ($10^{-2}$)} & 169 (0)                      & $1.3 \times 10^{-3}$                           & 2602       & 507        & 12         & $7.6\times 10^{-5}$ & $2.25\times 10^{-2}$ & 3.8 \\ \hline
\multicolumn{1}{|c|}{PIR. ($10^{-3}$)}  & 169 (0)                      & $1.3 \times 10^{-3}$                           & 2605       & 507        & 12         & $7\times 10^{-5}$ &  $2.31\times 10^{-2}$& 3.9 \\ \hline
\multicolumn{1}{|c|}{PIR. ($10^{-4}$)}  & 260 (1)                      & $8.1 \times 10^{-4}$                           & 3497       & 780        & 10         & $3.4\times 10^{-5}$ & $1.96\times 10^{-2}$& 5.1 \\ \hline
\multicolumn{1}{|c|}{PIR. ($10^{-5}$)}  & 800 (1)                      & $2.7 \times 10^{-4}$                           & 7234       & 2400       & 4          & $1\times 10^{-5}$ & $1.4\times 10^{-2}$& 11.3\\ \hline
\multicolumn{1}{|c|}{PIR. ($10^{-8}$)}  & 7318 (379)                   & $3.3 \times 10^{-5}$                            & 23472      & 22047      & 1          & $3.5\times 10^{-8}$ & $6\times 10^{-3}$& 42.2 \\ \hline
\end{tabular}
\vspace*{0.2cm}
\caption{Comparative analysis of explicit and PIROCK with varying tolerances ($tol$), showcasing the number of steps (including rejected steps), average timestep value ($\Delta t$), function evaluations for the $\Fdiff$ and $\Fadv$ terms, the maximum number of stages considered by the PIROCK method for the hyperdiffusive term, the L2 norm between the solution and the reference solution on the whole domain for the density variable $\rho$, the average CPU time per step for each case and the total computational time. Moreover, the computations were performed using a 2.4 GHz 8-Core Intel Core i9 processor.}
\label{tab:sodshock}
\end{table*}

\subsection{2.5 dimensional reconnection problem}

As previously outlined, our initial endeavor entails addressing a significant magnetic reconnection challenge involving four distinct species: $\{\text{H}, \text{H}^+, \text{He}, \text{He}^+\}$. Extending upon our previous study investigating magnetic reconnection with four species \citep[see][]{wargnier2022multifluid}, our objective is to showcase the applicability of the PIROCK method on the MFMS model in addressing complex solar physics phenomena involving multiple species, while maintaining computational efficiency. We compare with conventional methods such as explicit and splitting.

\subsubsection{Initial conditions}

We consider a reconnection problem with identical thermodynamic conditions but different box sizes and mixtures. The initial thermodynamic conditions for the 2.5D reconnection problem are based on those in \citet{wargnier2022multifluid} (Section 4.1), with \( T_0 = 16,000 \) K, \( n_0 = 4.8 \times 10^9 \) \(\text{cm}^{-3}\), and plasma beta \(\beta_p = 0.1\).

To compare the methods efficiently without performing an extensive analysis, we use a smaller computational domain. The smaller domain is necessary because the explicit and splitting methods are computationally very expensive, as shown below. The domain lengths in the \( y \) and \( z \) directions, denoted as \( L_y \) and \( L_z \), are 1.5 Mm and 3 Mm, respectively, with \( \Delta y = 100 \) km and \(\Delta z = 75 \) km, resulting in a grid of \( 150 \times 400 \) points and a magnetic field strength \( B_0 = 2 \) G. 


It is important to note that, unlike \citet{wargnier2022multifluid}, we have considered periodic boundary conditions along the \( z \) direction and open boundary conditions in the \( y \) direction, and the Hall term is taken into account. We compared the PIROCK method ($tol=10^{-2}$ and $10^{-5}$) with the splitting and explicit methods up to \( t=200\) s. 

\begin{table}[]
\begin{center}
\begin{tabular}{c|c|c|}
\cline{2-3}
& n{[}\%{]}            & Ionization level {[}\%{]} \\ \hline
\multicolumn{1}{|c|}{H}         & 34.6                 & \multirow{2}{*}{62.5}     \\ \cline{1-2}
\multicolumn{1}{|c|}{H$^+$}     & 57.6                 &                           \\ \hline
\multicolumn{1}{|c|}{He}        & 7.8                  & \multirow{2}{*}{0}        \\ \cline{1-2}
\multicolumn{1}{|c|}{He$^+$}    & $3.15\times 10^{-6}$ &                           \\ \hline
\end{tabular}
\vspace*{0.2cm}
\caption{Initial number densities in percentage in the 2.5 D reconnection problem. The initial number density of a given species $\alpha$ percentage are calculated as $n_{\alpha}\left[\%\right]=100\times n_{\alpha}/\sum_{\gamma}n_{\gamma}$. The initial ionization level of each species is shown in the right column.}
\label{tab:tablereco}
\end{center}
\end{table}

\begin{table*}
\centering
\begin{tabular}{c|c|c|c|c|c|c|c|c|}
\cline{2-9}
& No. steps & $\overline{\Delta t}$ & $\Fdiff$ evals & $\Fadv$ evals& $\Freac$ evals & $s_{\text{max}}$       & CPU time/step [s] & Tot. time [s] \\ \hline
\multicolumn{1}{|c|}{splitting}  & 821,974 & $5.7 \times 10^{-7}$ & 2,465,922 & 2,465,922 & 4,109,870 & 0 & $2\times 10^{-1}$ & $1.71\times 10^5$ \\ \hline
\multicolumn{1}{|c|}{explicit} & 1,103,531 & $4.59 \times 10^{-7}$& 3,310,593& 3,310,593& 3,310,593& 0& $9\times 10^{-2}$& $9.8\times 10^4$\\ \hline
\multicolumn{1}{|c|}{PIR. ($10^{-2}$)} & 3,306 & $1.486\times 10^{-4}$ & 38,516 & 9,918 & 18,701,425 & 8 & $1.45$ & $3.75\times 10^3$ \\ \hline
\multicolumn{1}{|c|}{PIR. ($10^{-5}$)} & 12,217 & $4.8\times 10^{-5}$ & 106,808 & 36,660 & 67,199,808 & 5 & $5.7\times 10^{-1}$ & $6.67\times 10^3$ \\ \hline
\end{tabular}
\vspace*{0.2cm}
\caption{Comparative analysis of all methods, including splitting, explicit and PIROCK with varying tolerances ($tol$), showcasing the number of steps, average timestep value ($\overline{\Delta t}$), function evaluations for the $\Fdiff$, $\Fadv$ and $\Freac$ terms, the maximum number of stages $s$, the average CPU time per step for each case, and the total computational time. The calculations were executed utilizing 28 Intel Broadwell processors integrated into the Pleiades cluster, consisting of the 14-core E5-2680v4 model operating at a clock speed of 2.4 GHz.}
\label{tab:ot}
\end{table*}

\begin{table*}[]
\centering
\begin{tabular}{c|c|c|c|c|}
\cline{2-5}
& $\text{H}$ & $\text{H}^+$ & $\text{He}$ & $\text{He}^+$ \\ \hline
\multicolumn{1}{|c|}{splitting}           &    $1.32\times 10^{-3}$        &      $1.9\times 10^{-3}$       &      $6.2\times 10^{-4}$      &        $2.3\times 10^{-6}$     \\ \hline
\multicolumn{1}{|c|}{explicit}           &    $6.89\times 10^{-4}$        &      $9\times 10^{-4}$       &      $7.13\times 10^{-4}$      &        $1.31\times 10^{-6}$     \\ \hline
\multicolumn{1}{|c|}{PIR. ($tol =10^{-2}$)} &    $5.44\times 10^{-4}$     &       $6.86\times 10^{-4}$    &      $5.8\times 10^{-4}$      &       $1.19\times 10^{-6}$     \\ \hline
\multicolumn{1}{|c|}{PIR. ($tol =10^{-5}$)}                  & $5.15\times 10^{-4}$      &  $6.49\times 10^{-4}$         & $5.67\times 10^{-4}$       &        $1.14\times 10^{-6}$         \\ \hline
\end{tabular}
\label{tab:L2}
\vspace*{0.2cm}
\caption{Comparison of L2 norm of the error of the species density relative to splitting, explicit and PIROCK (tol=$10^{-2}$ and $10^{-5}$) with respect to the reference solution.}
\end{table*}

\begin{figure*}[!tbh]
	\begin{center}
	   \hspace*{-0.4cm}
		\includegraphics[scale=0.4]{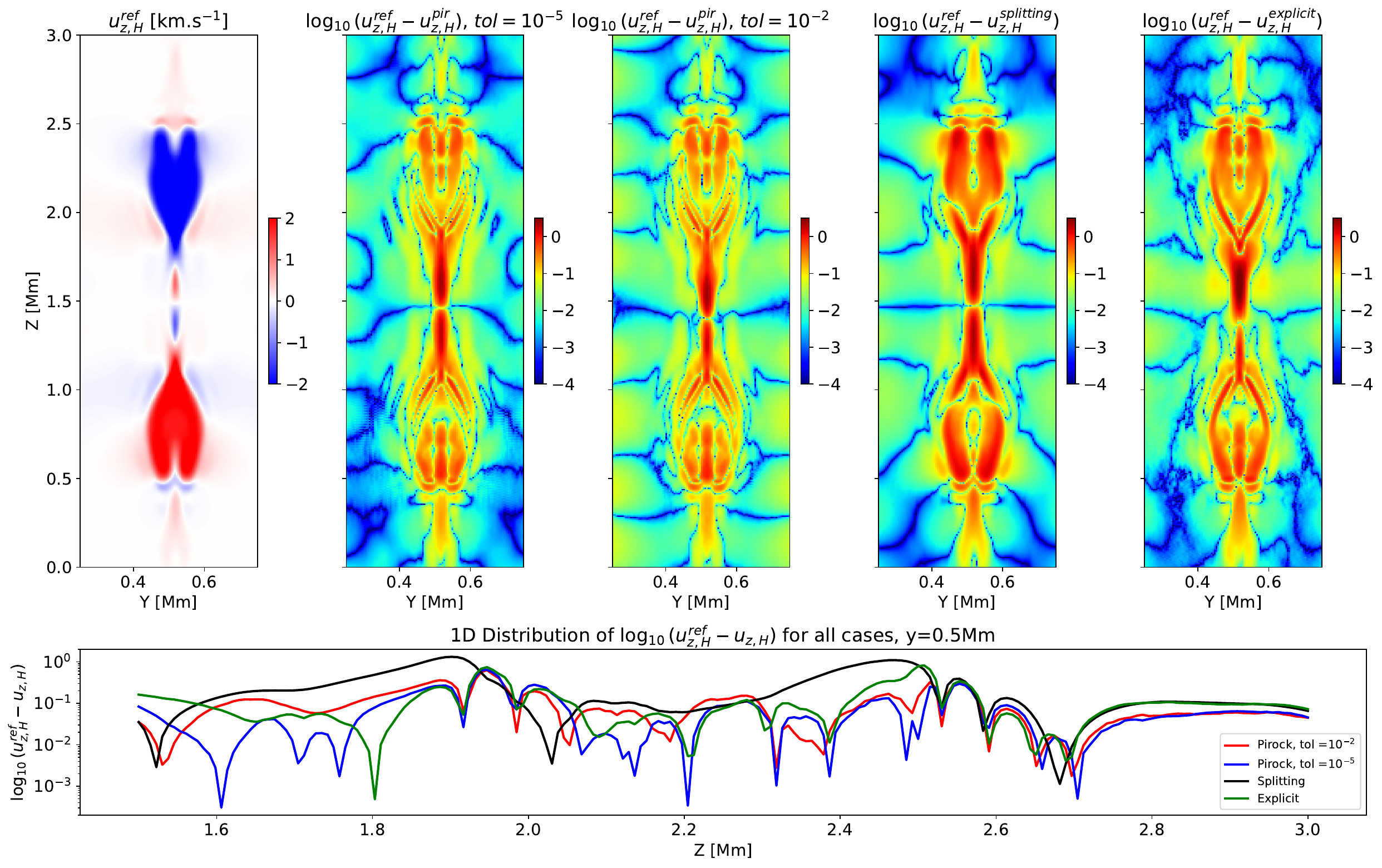}
		\vspace*{0.2cm}
		\caption{From left to right: 1) Distribution of the reference solution of the velocity $u_{z,H}$ in a restricted domain along the y direction ($y\in[0.2,0.8]$ Mm) in km.s$^{-1}$, 2D figures represent the logarithm of the difference between the reference and the numerical solution $\log_{10}(u^{ref}_{z,H}-u_{z,H})$ for PIROCK with tolerance $10^{-5}$, $10^{-2}$, splitting, and explicit. The last plot at the bottom represents the 1D distribution $\log_{10}(u^{ref}_{z,H}-u_{z,H})$ of all these cases at $y=0.5$ Mm and $z \in[1.5, 3]$ Mm at $t=50$~s.}
		\label{fig:comparison}
	\end{center}
\end{figure*}
\begin{figure*}[!tbh]
	\begin{center}
	   \hspace*{-0.4cm}
		\includegraphics[scale=0.4]{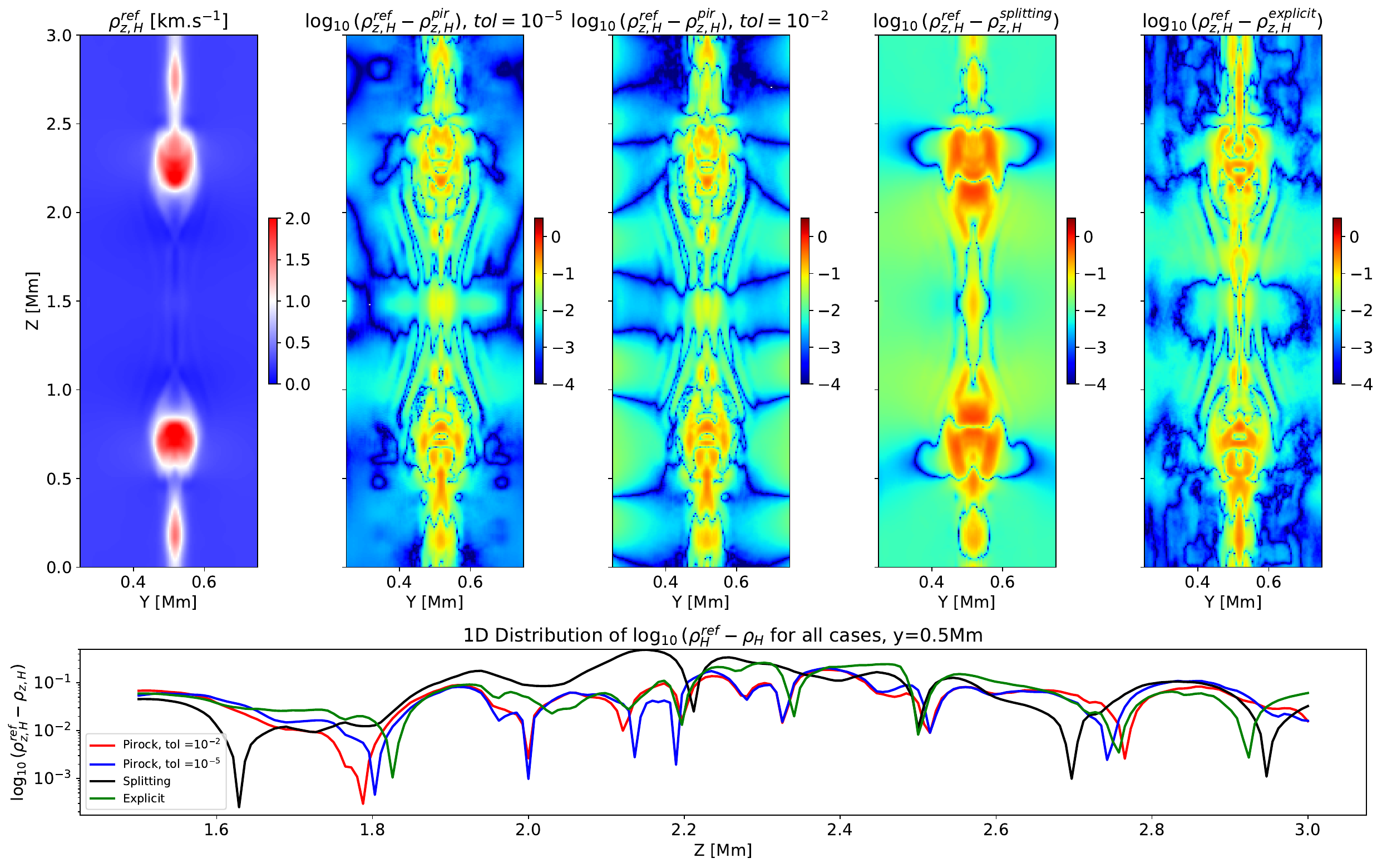}
		\vspace*{0.2cm}
		\caption{Same as Fig.~\ref{fig:comparison} but for $\rho_{H}$.}
		\label{fig:comparison2}
	\end{center}
\end{figure*}

\subsubsection{Comparative Analysis of Numerical Methods}

A reference solution was constructed using a fixed timestep of \(\Delta t = 10^{-8}\) s. The method used for this reference solution was a third-order Runge-Kutta scheme for the differential and advective terms ($\Fdiff$ and $\Fadv$), and an implicit scheme, similar to PIROCK, for the reactive term ($\Freac$).

In Table~\ref{tab:ot}, we compared the methods splitting, explicit, and PIROCK (with tolerances \(10^{-2}\) and \(10^{-5}\)) across various aspects such as the number of steps, average timestep value, number of function evaluations, maximum number of stages (\(s_{\text{max}}\)), average computational cost per step, and total computational cost of the simulation. The results clearly demonstrate the efficiency of the PIROCK method in terms of computational cost. Specifically, the number of steps required for PIROCK (3000 for \(tol=10^{-2}\) and 12000 for \(tol=10^{-5}\)) is significantly lower compared to explicit (1,000,000 steps) and splitting (800,000 steps). The average timestep (\(1.5 \times 10^{-4}\) and \(4.8 \times 10^{-5}\)) for PIROCK is at least three orders of magnitude larger than for explicit (\(4.59 \times 10^{-7}\)) and splitting (\(5.7 \times 10^{-7}\)). Even though the computational cost per step is higher, the total computational cost for PIROCK is dramatically lower than for the splitting and the explicit by almost two orders of magnitude.

In Table~\ref{tab:L2}, we present the L2 norm for all densities of the model compared to the reference solution. It is observed that for all species, the L2 norm is similar or slightly better for PIROCK. This indicates that for a much lower computational cost, we achieve nearly the same precision (or better) for the density. However, the improvement in precision between the tolerances \(10^{-2}\) and \(10^{-5}\) is not significant.

Figures~\ref{fig:comparison} and \ref{fig:comparison2} illustrate the final time distribution of the velocity \(u_{z,H}\) and the density \(\rho_H\), along with a 2D comparison of the difference between the reference and the solution obtained for each method. Focusing on Fig.~\ref{fig:comparison}, we observe that the error is primarily localized in the current sheet, the outflow regions of the reconnection, and strong discontinuities (near plasmoids) and shocks. However, PIROCK appears to capture the solution better in these shock regions, as shown by the 1D distribution (bottom panel). It is also evident that splitting is quite imprecise compared to the other methods. This imprecision arises because we use a first-order Lie splitting, which introduces a first-order temporal error, making the method much less accurate than the explicit and PIROCK methods, which are of orders 3 and 2, respectively. Similar results are obtained for all other variables.

PIROCK proves to be highly efficient regarding both computational cost and accuracy for all variables. The number of steps is significantly reduced, allowing for a much larger timestep and a drastically lower computational cost than traditional splitting and explicit methods.

\subsubsection{Fluxes calculation and chemical fractionation}

In this section, we demonstrate the capabilities of PIROCK in efficiently solving multi-fluid reconnection problems. As an example, we quantify the chemical fractionation between hydrogen and helium during magnetic reconnection. We conduct a study similar to \citet{wargnier2022multifluid} but with significant differences. In our case, the flux calculations in the outflow ($z$) direction are performed over a much larger domain, which promotes increased plasmoid formation. Additionally, periodic boundaries are used instead of open boundaries throughout the entire setup.

Additionally, the grid size and resolution of the current sheet are different; we note here that the spatial resolution is 150x400 grid points, with $\Delta y = 100$~km and $\Delta z = 75$~km, whereas in \citet{wargnier2022multifluid} it was 800x800 with $\Delta y = \Delta z = 5.33$~km. Moreover, although the domain is periodic in the $z$-direction, we have verified that the boundary conditions do not influence these flux calculations.

To characterize the chemical fractionation in the outflows, similar to \citet{wargnier2022multifluid}, the fluxes of helium and hydrogen species have been calculated and compared at the exhaust of the reconnection event until $t=180$~s. These calculations are conducted at fixed heights of $z_i = 1.9$~Mm and $1.7$~Mm. For any species $\alpha$, the corresponding flux in number density is calculated as

\begin{equation} \phi(\alpha) = \Delta t \Delta x \int_{l_y} \left[ u_{z,\alpha} n_{\alpha} \right](y, z = z_i) dy, \end{equation}
where $\Delta x$ is the grid size in the $x$ direction (set to unity), and $l_y$ is the length in the $y$ direction limited to the outflow region. The outflow regions are defined as the FWHM of the maximum outflow velocity at $z = z_i$. This length is obtained by locating the maximum outflow velocity at $z = z_i$, and the termination of the exhaust is approximated at $90\%$ of this maximum in the $y$ direction to estimate $l_y$.

\begin{figure}[!tbh]
    \centering
    \includegraphics[scale=0.38]{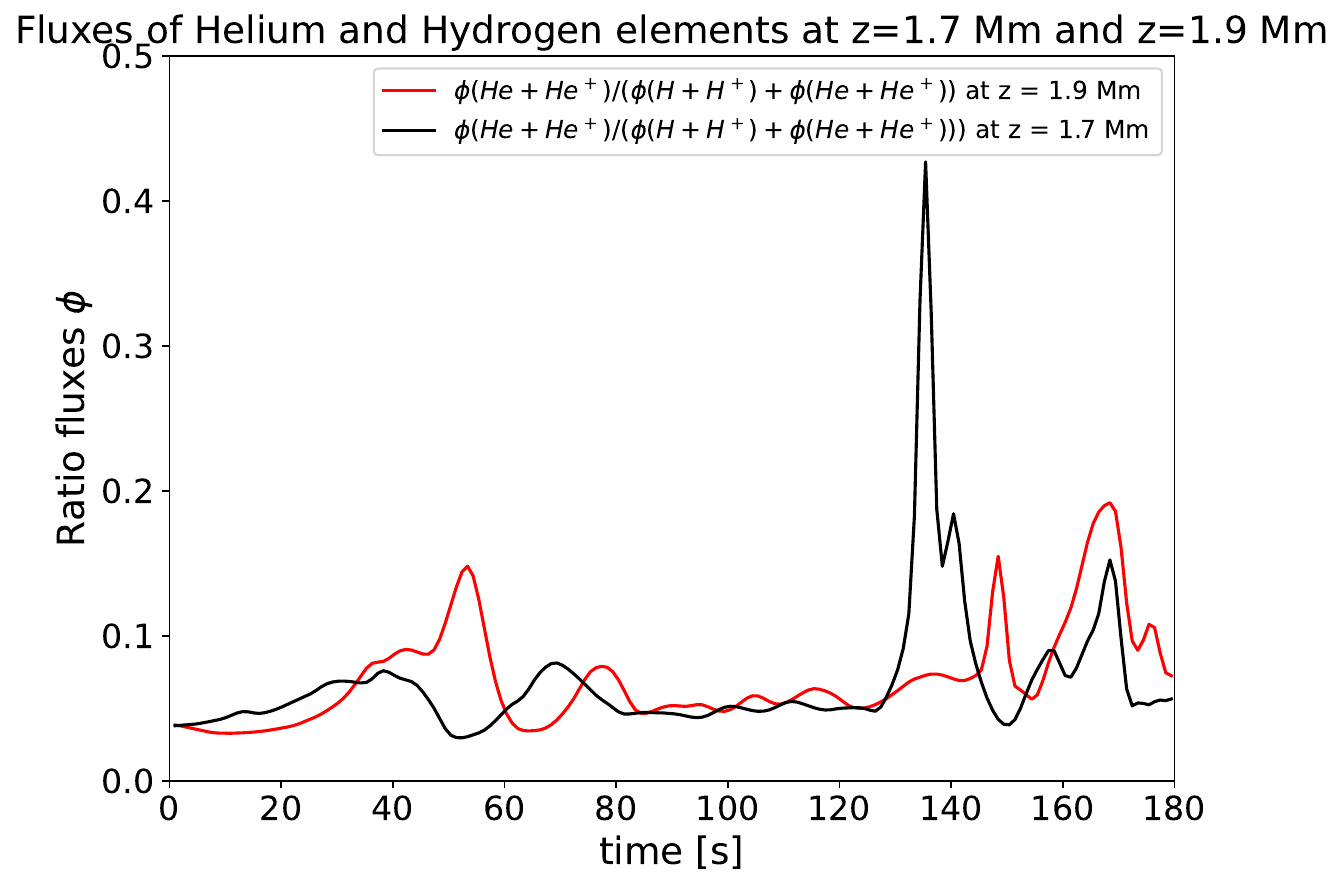}
    \caption{Time evolution of the flux ratio of helium species relative to the total flux of all species, $\phi(\text{He} + \text{He}^+)/(\phi(\text{He} + \text{He}^+) + \phi(\text{H} + \text{H}^+))$, at two different locations: $z=1.7$~Mm (black line) and $z=1.9$~Mm (red line).}
    \label{fig:flux}
\end{figure}

In Fig.~\ref{fig:flux}, which shows the time evolution of the flux ratio $\phi(\text{He} + \text{He}^+)/(\phi(\text{He} + \text{He}^+) + \phi(\text{H} + \text{H}^+))$ at different heights, $z=1.7$~Mm and $z=1.9$~Mm, we observe significant changes in the flux distribution as the reconnection evolves. Focusing on the black curve at $z=1.7$~Mm, as turbulence develops with the formation of plasmoids and shocks, the helium flux increases significantly relative to the total flux. This indicates that during the passage of shocks, a higher fraction of helium is produced than hydrogen, with the flux ratio reaching up to $42\%$, as observed around $t \approx 140$~s during a plasmoid event.

As the shock propagates and reaches $z=1.9$~Mm (red curve), a similar increase in helium flux can be seen, corresponding to earlier events at $z=1.7$~Mm. For example, the rise in flux at $t \approx 140$~s in the black curve corresponds to a peak at $t = 150$~s in the red curve. This suggests the presence of multiple shocks and plasmoids propagating and causing peaks in the flux ratio. The helium fraction increases significantly during turbulent reconnection. The mechanism explaining this increase is similar to what is described in \citet{wargnier2022multifluid}. Briefly, during the laminar phase, the ionization fraction of hydrogen is higher, and hydrogen is slowly expelled, with relatively small outflows. However, in the turbulent phase, characterized by stronger flows and the formation of plasmoids, the number of helium elements increases, and helium is ionized due to the temperature rise, resulting in an enrichment of helium in the outflows. The difference, as explained earlier, lies in the resolution, the number of grid points, the domain size, and the evolution of the magnetic field topology due to the periodic boundary conditions, which are distinct between the two cases.

\section{Conclusion}~\label{sec:realscenar}

This study concentrated on devising a novel and efficient temporal integration method tailored to the multi-fluid MHD model. The latter has been introduced and reformulated as a set of ordinary differential equations, elucidating the diffusive, convective, and stiff source terms within the model. However, we point out that the outlined numerical strategy holds applicability to any single or multi-fluid MHD system provided that we can reformulate the system of equations as a stiff reaction-diffusive-convective set of equations. Notably, the numerical strategy remains agnostic to the specifics of spatial discretization.

We presented an overview of Runge-Kutta methods commonly used for integrating diffusive, stiff reactive, and advective terms, aiming to identify the most suitable numerical time integration method. In solar and stellar physics, while various methods excel in integrating stiff reactive or advective terms, diffusive terms pose a primary challenge. In this work, we introduce, for the first time in the context of solar and stellar physics, the well-known ROCK2 method. This approach is particularly attractive due to its explicit, second-order, and competitive nature with implicit schemes. Another challenge is integrating diffusion-advection and reactive terms simultaneously. Traditional approaches like time operator splitting have been considered, albeit with known splitting errors and stability issues for multiple stiff terms. We focus on the PIROCK method, developed by \citet{ABDULLE2013869}, an implicit-explicit partitioned Runge-Kutta method, for integrating diffusive, stiff reactive, and advective terms altogether. PIROCK combines ROCK2 for diffusion, SDIRK for stiff reactive terms, and a third-order Heun method for advective terms, incorporating an adaptive timestep strategy and error calculations using methods introduced by \citet{Hairer1993}. It is important to note that the artificial hyperdiffusive terms in the model are integrated using the ROCK2 method simultaneously with all other physical diffusion terms.

We conducted a comparative analysis of the PIROCK method against two other approaches: a purely explicit method using a 3rd-order Runge-Kutta scheme with an explicit integration of reactive and diffusive terms and a splitting approach employing Lie Splitting identically to \citet{wargnier2022multifluid}. Focusing initially on a single-fluid MHD Sod-Shock tube problem, we assessed the explicit and PIROCK methods for varying tolerances. Results demonstrate that user-defined tolerances effectively regulate solution quality and computational cost, as expected. By choosing to integrate hyperdiffusive terms within the ROCK2 part of PIROCK, it becomes possible to control the quality of the numerical solution, especially in shocks and discontinuities, using the properties of embedded methods, which is not achievable with standard explicit methods due to the time step being always limited by the CFL constraint of hyperdiffusive terms. In this context, it is feasible to reduce the computational cost from 13.3 s (2203 steps) to just 3.8 s (160 steps) (for $tol =10^{-2}$) with a loss of accuracy from $2\times 10^{-6}$ to $7.6\times 10^{-5}$ as detailed in Table \ref{tab:sodshock}. However, we emphasize that the solution quality remains highly satisfactory, and this loss of precision is simply associated with increased diffusion and regularization around discontinuities and shocks. In this respect, PIROCK demonstrates greater flexibility (regarding computational cost versus accuracy) compared to the traditional explicit approach that involves hyperdiffusive terms.

Next, we focused on a 2D multi-fluid reconnection problem involving multiple species $\{\text{H},\, \text{H}^+,\text{He},\,\text{He}^+\}$ with the complete model including all terms. Results show that PIROCK outperforms splitting and explicit for all variables compared to our constructed reference solution. Regarding computational cost only, PIROCK outperforms splitting and is more than one, almost two, orders of magnitude faster for the highest tolerances. PIROCK has a slightly higher per-step computational cost due to the high number of source term evaluations $\Freac$ and the chosen number of stages $s$ for the integration of diffusive terms with ROCK2. Nevertheless, at a tolerance of $10^{-2}$, PIROCK requires only 3,306 steps to reach the final time $t=t_f=50 s$, whereas splitting requires 821,974, hence a number of timesteps reduced by a factor $250$. Concerning accuracy, it is clear that PIROCK outperforms the standard splitting approach, and the latter shows an excess of diffusion, especially close to shocks and discontinuities. The excessive diffusion in the splitting approach is attributed to first-order temporal errors due to the Lie splitting error, while PIROCK is a second-order scheme.

Finally, thanks to this new numerical method, we are now able to tackle complex multi-fluid problems that were previously unsolvable in reasonable computational time. We have demonstrated the ability of our numerical model to address chemical fractionation between hydrogen and helium species in the context of magnetic reconnection. Our results highlight the decoupling of species when the reconnection regime becomes turbulent similarly, as in \citet{wargnier2022multifluid}. These cases show the potential of the numerical model for future multi-fluid applications such as the First-Ionization-Potential (FIP) effect, which required more fluids (high and low FIP elements) and is now feasible thanks to PIROCK. The latter will be investigated in a follow-up paper.

In summary, the PIROCK method exhibits remarkable superiority over conventional approaches like standard splitting methods and standard explicit methods in terms of accuracy, stability, and computational efficiency, regardless of tolerance levels. This superiority is evident across various scenarios, including typical two-dimensional solar physics situations and hyperbolic Sod shock problems, covering both the multi-fluid system of equations \eqref{eq:system} and the single-fluid approach. These findings underscore the method's exceptional performance compared to established classical time integration methods.

\begin{acknowledgements}
The authors would like to thank Ernst Hairer for helpful discussions. This project received partial financial support from NASA grants 80NSSC21K0737, 80NSSC21K1684, and contracts NNG09FA40C (IRIS) and 80GSFC21C0011 (MUSE). The simulations have been run in the Pleiades cluster through the computing projects s1061, s2601, and s2967 from the High-End Computing (HEC) division of NASA.
The second author was partially supported by the Swiss National Science Foundation, projects No 200020\_214819, and No. 200020\_192129.
\end{acknowledgements}

%
\bibliographystyle{aa} 
\bibliography{collectionbib} 
%
\begin{appendix}

\section{ROCK2 algorithm for diffusive terms}
\label{app:rock2}

For a given number of stage $s>1$ and an initial condition $Y_0$, the ROCK2 time integration method applies to the problem $\frac{d\consvar(t)}{dt} = \Fdiff\left(\consvar(t)\right)$ with initial condition $Y_0$, reads

 \begin{equation}
	\begin{aligned}
	\consvar_1 &= \consvar_0 + \alpha\mu_1\Delta t\Fdiff(\consvar_0)  \\ 
    \consvar_j &= \alpha\mu_j\Delta t\Fdiff\left(\consvar_{j-1}\right)-\nu_j \consvar_{j-1} - \kappa_j \consvar_{j-2},\\
    \consvar_{s-1} &= \consvar_{s-2} +\sigma_{s}\Delta t\Fdiff\left(\consvar_{s-2}\right)\\
     \consvar_{s} &= \consvar_{s-1}+\sigma_{s}\Delta t\Fdiff\left(\consvar_{s-1}\right)\\
     \consvar_{n} &= \consvar_{s} - \sigma_{s}\Delta t(1-\tfrac{\tau_{s}}{\sigma^2_{s}}) \Delta \Fdiff^s \\
   \Delta \Fdiff^s&=\left[\Fdiff\left(\consvar_{s-1}\right) - \Fdiff\left(\consvar_{s-2}\right) \right]
		\end{aligned}
  \tag{ID}
\label{eq:diff}
	\end{equation}
where $j=2,\ldots,s-2,$  $\mu_j, \nu_j$ and $\kappa_j$ are obtained from the Chebyshev recurrence relation of the orthogonal Chebyshev polynomials (as described in Eqs.~24-25 from \citet{ROCK2_2001} and defined as $\{P_j\}_{j\geq 0}$ in \citet{ABDULLE2013869}), and $\consvar_{n}$ is the vector conservative variables after integration. Note that the Chebyshev polynomials $\{P_j\}_{j\geq 0}$ are the stability functions of each internal stage of the ROCK2 method (Eq.~\eqref{eq:diff}). 
The orininal ROCK2 methods uses $\alpha=1$ while we use a parameter $\alpha>1$ in PIROCK.
The values of the coefficients $\sigma_{s}$ and $\tau_{s}$ depend on the total number of stage $s$ and are chosen such that they satisfy second order accuracy of the method for any $s$.

 \section{2-stage second-order Singly Diagonally Implicit Runge-Kutta (SDIRK) method for source terms}
 \label{app:DIRK}
The 2-stage second-order and L-stable Singly Diagonally Implicit Runge-Kutta (SDIRK) method applies to the problem $\frac{d\consvar(t)}{dt} = \Freac\left(\consvar(t)\right)$ with initial condition $Y_0$, reads
 
  \begin{equation}
	\begin{aligned}
	\consvar_1 &= \consvar_0 + \gamma\Delta t\Freac(\consvar_1)  \\ 
    \consvar_2 &= \consvar_0 +(1-2\gamma) \Delta t\Freac(\consvar_1)+\gamma\Delta t\Freac(\consvar_2) \\
    \consvar_{n} &= \consvar_0 + \frac{\Delta t}{2}\Freac(\consvar_1) + \frac{\Delta t}{2}\Freac(\consvar_2).
		\end{aligned}
  \tag{IR}
\label{eq:source}
	\end{equation}
 where $\gamma = 1-\sqrt{2}/2$ and $\consvar_{n}$ is the vector of conservative variables after integration.

\section{PIROCK algorithm}
\label{app:pirock}
\subsection{Scenario with Dominant Diffusive Processes ($s>1$)}
For thoroughness, we present the formulation of the PIROCK method, when $s>1$ with initial condition $Y_0$, for the integration of the complete system in Eq.~\eqref{eq:system}, as detailed in \cite{ABDULLE2013869},

	\begin{equation}
	\begin{aligned}
    &\text{Diffusion stabilization procedure}\\
	&\consvar_1 = \consvar_0 + \alpha\mu_1\Delta t\Fdiff(\consvar_0), \\
     &\consvar_j = \alpha\mu_j\Delta t\Fdiff\left(\consvar_{j-1}\right)-\nu_j \consvar_{j-1} - \kappa_j \consvar_{j-2}, \quad  j=2,\ldots,s \\
     &\text{Finishing procedure for diffusion (additional two steps)}\\
     &\widetilde{\consvar}_{s-1} = \consvar_{s-2} +\sigma_{s}\Delta t\Fdiff\left(\consvar_{s-2}\right), \\
     &\widetilde{\consvar}_{s} = \widetilde{\consvar}_{s-1} +\sigma_{s}\Delta t\Fdiff\left(\widetilde{\consvar}_{s-1}\right), \\
     &\text{Integration of advective-reaction terms and coupling with diffusion} \\
     &\consvar_{s+1} =  \consvar_{s} + \gamma\Delta t\Freac\left(\consvar_{s+1}\right), \\
     &\consvar_{s+2} = \consvar_{s} + \Delta t(1-2\gamma) \Freac\left(\consvar_{s+1}\right)+\Delta t\gamma\Freac\left(\consvar_{s+2}\right) + \Delta t \Fadv\left(\consvar_{s+1}\right), \\
      &\consvar_{s+3} = \consvar_{s} + \Delta t(1-2\gamma)\Fadv\left(\consvar_{s+1}\right) + \Delta t(1-\gamma)\Freac\left(\consvar_{s+1}\right), \\
       &\consvar_{s+4} = \consvar_{s} + \frac{\Delta t}{3}\Fadv\left(\consvar_{s+1}\right), \\
       &\consvar_{s+5} = \consvar_{s} + \frac{2\Delta t}{3}\bold J^{-1}_{\text{R}} \Fadv\left(\consvar_{s+4}\right) + \Delta t\left(\frac{2}{3}-\gamma\right)\Freac\left(\consvar_{s+1}\right) \\ &\quad\quad\quad\quad+\frac{\Delta t 2\gamma}{3}\Freac\left(\consvar_{s+2}\right)\\
       &\text{Final integration from $\consvar_0$ to $\consvar_{n}$} \\
       &\consvar_{n} = \widetilde{\consvar}_{s} -\sigma_{s}\Delta t(1-\tfrac{\tau_{s}}{\sigma^2_{s}})\left[\Fdiff\left(\widetilde{\consvar}_{s-1}\right) - \Fdiff\left(\consvar_{s-2}\right) \right]\\ &\quad \quad+\frac{\Delta t}{4}\Fadv\left(\consvar_{s+1}\right) +  \frac{3\Delta t}{4}\Fadv\left(\consvar_{s+5}\right)  \\&\quad\quad\,+ \frac{\Delta t}{2}\Freac\left(\consvar_{s+1}\right)+ \frac{\Delta t}{2}\Freac\left(\consvar_{s+2}\right)\\&\quad\quad\, + \frac{\Delta t \bold J^{-1}_{\text{R}}}{2-4\gamma}\left[\Fdiff\left(\consvar_{s+3}\right) - \Fdiff\left(\consvar_{s+1}\right) \right]
	\end{aligned}
 \tag{P}
\label{eq:pirock}
	\end{equation}
 where $\bold J_{\text{R}} = \identity-\gamma\Delta t\tfrac{\partial\Freac}{\partial\consvar}(\consvar_s)$, $\gamma$, $\tau_{s}$, $\sigma_{s}$, $\nu_j$,  $\kappa_j$ and $\mu_j$ are coefficients that have been described in Sections~\ref{app:rock2} and \ref{app:DIRK}, $\alpha = 1/(2P'_{s}(0))$ (chosen such that the internal stage $\consvar_{s}$ integrates the diffusion part alone for half stepsize $\alpha P'_{s}(0) \Delta t = \Delta t/2$) and $\consvar_{n}$ is the vector of conservative variable after integration. Note that the PIROCK algorithm \ref{eq:pirock} is identical to the one presented in \citet{ABDULLE2013869} where we choose the particular case where $\alpha>1$  and stage parameter $s$ given in Eq.~\eqref{eq:defs0.43} (due to the stability domain size $0.43 s^2$ along the negative real axis for the diffusion terms). We emphasize the negligible computational expense incurred in computing $\bold J^{-1}_{\text{R}}$, where $\bold J^{-1}_{\text{R}}$ represents the inverse of the Jacobian matrix tied to the source terms $\Freac$. This efficiency stems from the fact that the LU factorization of the Jacobian $\bold J_{\text{R}}$ is readily obtainable during the computation of the implicit stages of $\Freac$.

\subsection{Scenario where $s=1$}

In contrast to the preceding subsection, we now examine the PIROCK case with $s=1$. This corresponds to a scenario where the dominant advective process within the same system (S) is considered. For an initial condition $Y_0$, the algorithm in this context is as follows:

	\begin{equation}
	\begin{aligned}
     &\consvar_{1} =  \consvar_{0} + \gamma\Delta t\Freac\left(\consvar_{1}\right), \\
     &\consvar_{2} = \consvar_{0} + \Delta t(1-2\gamma) \Freac\left(\consvar_{1}\right)+\Delta t\gamma\Freac\left(\consvar_{2}\right)\\&\quad \quad+ \Delta t \Fadv\left(\consvar_{1}\right) + \Delta t \Fdiff\left(\consvar_{1}\right), \\
       &\consvar_{3} = \consvar_{0} + \frac{\Delta t}{3}\Fadv\left(\consvar_{1}\right) + \frac{\Delta t}{3}\Fdiff\left(\consvar_{1}\right), \\
       &\consvar_{n} = \consvar_{0} + \frac{2\Delta t}{3}\bold J^{-1}_{\text{R}} \left(\Fadv\left(\consvar_{3}\right)+\Fdiff\left(\consvar_{3}\right)\right) \\ & \quad\quad+ \Delta t\left(\frac{2}{3}-\gamma\right)\Freac\left(\consvar_{1}\right) +\frac{\Delta t 2\gamma}{3}\Freac\left(\consvar_{2}\right)
	\end{aligned}
 \tag{P2}
\label{eq:pirock2}
	\end{equation}
where $\gamma = 1-\sqrt{2}/2$ and $\consvar_{n}$ is the vector of conservative  variables after integration.

\end{appendix}

\end{document}